\documentclass[nofootinbib,aps,prb,twocolumn,superscriptaddress,floatfix]{revtex4-2}
\usepackage[english]{babel}
\usepackage[utf8]{inputenc}
\usepackage[T2A]{fontenc}
\usepackage{hyperref, xcolor, graphicx}
\usepackage{amsfonts, amssymb, amsmath, mathtools}
\usepackage{subcaption}
\usepackage{siunitx}
\usepackage{tikz}
\usepackage{calc}
\usepackage{float}
\usepackage{ragged2e}
\usepackage{soul}
\usepackage{dcolumn}
\usepackage[normalem]{ulem}
\usepackage{xcolor,colortbl}
\usepackage{array, makecell} 
\definecolor{Gray}{gray}{0.85}

\hypersetup{
	colorlinks=true,
	linkcolor=black,
	citecolor=blue,   
	urlcolor=blue,
}

\makeatletter
\def\maketitle{
	\@author@finish
	\title@column\titleblock@produce
	\suppressfloats[t]}
\makeatother

\begin{document}

	\title{Crossover of magnetodipole and electrodipole contributions in the broadband magneto-optics of bismuth-substituted iron garnets}
	
	\author{N.I.\ Gribova}
	\email{gribova.ni@phystech.edu}
	\affiliation{Abrikosov Center for Theoretical Physics, Moscow Institute of Physics and Technology, 9, Institutsky per, Dolgoprudny 141701, Russia}
	\date{\today}
	
	\begin{abstract}
				This work presents a comprehensive macroscopic electrodynamic framework isolating the magnetodipole (MD) and electrodipole (ED) contributions to the magneto-optical effects in bismuth-substituted iron garnet. By rigorously incorporating the dispersive magnetic permeability tensor, the dominance of the MD mechanism in the terahertz (THz) gap is demonstrated. In the mid-infrared (MIR) to near-infrared (NIR) domain, a fundamental spectral crossover is revealed where the magnitudes of the MD and ED contributions exactly equalize, invalidating the widespread assumption that magnetic contributions vanish uniformly at infrared frequencies. Furthermore, it is shown that interfacial phase shifts uniquely dictate how these competing mechanisms project onto observable rotation and ellipticity, yielding an inverted spectral hierarchy between transmission (Faraday) and reflection (polar and longitudinal Kerr) geometries. Most notably, the transverse  Kerr effect for s-polarized light is established to be exclusively driven by the off-diagonal magnetic permeability. While strictly zero under the conventional ED approximation, the s-polarized transverse Kerr effect emerges as a robust, free of the gyroelectric contribution probe of MD activity in the THz and MIR--NIR regimes, yielding a THz response two orders of magnitude larger than its p-polarized counterpart.
	\end{abstract}
	
	\maketitle
	

		\section*{Introduction}
		The terahertz (THz) frequency range has emerged as an important regime for studying ultrafast spin dynamics, coherent magnetic excitations, and spintronic phenomena, with extensive experimental and theoretical work demonstrating direct THz control of magnetic order and spin currents \cite{walowski2016perspective,blank2021thz,blank2023rectification}. Its relevance to magneto-optics is particularly strong because the THz field can couple directly to collective magnetic modes while optical measurements provide a sensitive probe of their dynamics. Ferrimagnetic iron garnets are especially promising in this respect because their two-sublattice magnetic structure supports both low-frequency acoustic and high-frequency optical spin excitations \cite{princep2017full,shamoto2018neutron}. Bismuth substitution further enhances and tailors the magneto-optical response of iron garnets, including the Faraday rotation and their spectroscopic properties \cite{takeuchi1973faraday,wittekoek1975magnetooptic,hansen1984bismuth,hansen1984review,ozgur2009microwave}. Most importantly, the coupled dynamics of the two Fe sublattices gives rise to the Kaplan--Kittel exchange resonance predicted for ferrimagnets \cite{kaplan1953exchange}, observed in the infrared through exchange-related magneto-optical effects \cite{krinchik1962exchange}, and directly identified in YIG at approximately 7.8~THz \cite{hsu2020observation}. Direct THz excitation of exchange modes has also been demonstrated in iron garnets using ultrafast THz magnetic fields \cite{blank2021thz}. These properties make bismuth-substituted ferrimagnetic garnets a natural platform for investigating the THz magneto-optical response associated with high-frequency ferrimagnetic dynamics.

At first sight, however, magneto-optical effects in the THz range are expected to be weak because the conventional electronic electric-dipole (ED) contribution rapidly decreases as the photon energy moves away from the electronic resonances that dominate the visible and near-infrared spectrum \cite{pershan1967magnetooptical,freiser1968survey,zvezdin1997modern,wettling1976magnetooptics}. This picture suggests that the magneto-optical response should become negligible at sufficiently low photon energies, but the experimental history of iron garnets indicates otherwise. As early as 1960--1962, Krinchik and Chetkin observed a pronounced Faraday response in YIG and other garnets throughout the infrared spectral range \cite{krinchik1960faraday,krinchik1961infrared,krinchik1962exchange}, and their analysis identified magnetic-dipole (MD) and exchange-related contributions to the infrared magneto-optical response \cite{krinchik1962exchange}. Later measurements of rare-earth iron garnets explicitly separated ED and MD contributions to the Faraday effect \cite{crossley1969faraday} and established their strong spectral dependence in the 1.1--1.7~$\mu$m range \cite{booth1984garnet}. More recently, Gribova \textit{et al.} demonstrated that the MD contribution remains essential in the NIR and can dominate the inverse Faraday effect in iron garnet films \cite{gribova2024inverse}; their analysis showed that the NIR response cannot be described consistently without accounting for both electric- and magnetic-dipole mechanisms. Thus, the apparent suppression of magneto-optics at long wavelengths does not imply the disappearance of the magnetic response; rather, it signals a crossover from an ED-dominated regime toward a regime in which the MD contribution becomes increasingly important.

This crossover requires an extension of the conventional magneto-optical description. Magneto-optical effects are usually formulated through the antisymmetric, off-diagonal components of the dielectric-permittivity tensor $\hat{\varepsilon}$ \cite{pershan1967magnetooptical,freiser1968survey,zvezdin1997modern}, whereas a complete electrodynamic treatment of a magnetically active medium must also include the magnetic-permeability tensor $\hat{\mu}$ \cite{krinchik1959tensors,wangness1954susceptibility,atkinson1992sign}. In ferrimagnetic garnets, the off-diagonal component $\mu_{xy}$ originates from the dynamic response of the coupled magnetic sublattices and therefore contains the two characteristic ferrimagnetic branches: the low-frequency acoustic mode and the high-frequency optical, or Kaplan--Kittel, mode \cite{wangness1954susceptibility,princep2017full,shamoto2018neutron,hsu2020observation}. The importance of this magnetic contribution is supported experimentally by the correlation between the oscillator strength of Kaplan--Kittel exchange modes and the magnetic permeability in far-infrared measurements of Dy$_3$Fe$_5$O$_{12}$ \cite{kang2012farinfrared}. At the same time, the optical response retains contributions from electronic transitions and Fe--O charge-transfer processes \cite{moskvin2022charge}, so that the observed magneto-optical signal in the NIR and THz regimes results from the interplay of ED and MD mechanisms. Consequently, the long-wavelength response of ferrimagnetic garnets cannot be described reliably by $\hat{\epsilon}$ alone; the dispersive off-diagonal permeability $\mu_{xy}$ must be treated on the same footing, particularly in the spectral region where the magnetic contribution becomes dominant \cite{gribova2024inverse,kang2012farinfrared}.

The role of the off-diagonal magnetic permeability in optical magneto-optics has been demonstrated experimentally in a magneto-plasmonic metamaterial, where a nonzero $\mu_{xy}$ was shown to generate an $s$-polarized transverse Kerr response that is absent for a purely gyroelectric medium \cite{yang2022gyromagnetic}. In the present work, the complementary problem of the intrinsic gyromagnetic response of a ferrimagnetic garnet is addressed, where $\mu_{xy}(\omega)$ arises from the coupled dynamics of the two magnetic sublattices and contains both the acoustic and Kaplan--Kittel modes.

In this work, the broadband magneto-optical response of bismuth-substituted ferrimagnetic garnets is established by explicitly separating the ED and MD contributions and following their evolution from the NIR into the THz range. It is demonstrated that the MD mechanism becomes the dominant contribution in the THz gap and determine the spectral condition at which the MD and ED contributions are exactly equal in the MIR--NIR. By extending the tensorial description to transmission and reflection geometries, it is shown that the Faraday and Cotton--Mouton effects provide sensitive optical probes of the Kaplan--Kittel exchange resonance. Further , it is demonstrated that interfacial phase shifts determine the spectral hierarchy of the different magneto-optical geometries: the transmission (Faraday) and polar Kerr responses exhibit opposite spectral behavior in the MIR/NIR, whereas the $s$-polarized transverse Kerr effect decouples from the ED background and, in both the THz and MIR/NIR regimes, is governed exclusively by the MD contribution through $\mu_{xy}$. In the THz gap, this $s$-polarized transverse Kerr signal is two orders of magnitude larger than the corresponding $p$-polarized response, providing a background-free optical signature of MD spin dynamics. Finally, the longitudinal Kerr geometry exhibits an additive coupling of the ED and MD contributions, producing a pronounced ellipticity extremum, where an equal-magnitude crossover between the MD-reference and ED-associated responses. These results demonstrate that magnetic contributions cannot be arbitrarily neglected at high frequencies and establish a unified framework for identifying and probing ferrimagnetic exchange dynamics through THz and NIR magneto-optics.

	\section*{Results}

Light propagation within a continuous medium is governed by Maxwell's equations, where the electromagnetic response of the material is characterized by the dielectric permittivity tensor $\hat{\varepsilon}$ and the magnetic permeability tensor $\hat{\mu}$. Throughout this work, the magnetization $\mathbf{M}$ is assumed to be oriented along the $z$-axis, inducing uniaxial anisotropy. Under these conditions, the constitutive tensors take the following form:
\begin{equation}
	\hat{\varepsilon} = 
	\begin{pmatrix}
		\varepsilon_{xx} & \varepsilon_{xy} & 0 \\
		-\varepsilon_{xy} & \varepsilon_{xx} & 0 \\
		0 & 0 & \varepsilon_{zz} 
	\end{pmatrix},\quad
	\hat{\mu} = 
	\begin{pmatrix}
		\mu_{xx} & \mu_{xy} & 0 \\
		-\mu_{xy} & \mu_{xx} & 0 \\
		0 & 0 & \mu_{zz} 
	\end{pmatrix}, \label{tensors}
\end{equation}
where the tensor elements are complex-valued quantities defined as $\varepsilon_{ij} = \varepsilon_{ij}^{\prime} + i \varepsilon_{ij}^{\prime \prime}$ and $\mu_{ij} = \mu_{ij}^{\prime} + i \mu_{ij}^{\prime \prime}$.

In the present work, both the dielectric permittivity and magnetic permeability are treated as frequency-dependent tensors throughout the MIR--NIR and THz spectral ranges. The dielectric tensor is reconstructed from the experimentally determined visible-to-NIR spectra by fitting the diagonal and off-diagonal components with a resonant model and subsequently extrapolating the fitted low-energy tails into the spectral region where direct experimental data for BiYIG are unavailable (Appendix~\ref{tensor_eps}). This procedure preserves the resonant spectral structure of $\hat{\varepsilon}$ while providing a physically consistent representation of its low-frequency response. All magneto-optical spectra of Bi$_x$Y$_{3-x}$Fe$_5$O$_{12}$ shown below correspond to the bismuth consentration \(x=0.2\), for which the dielectric tensor was reconstructed as described in Appendix Section~\ref{tensor_eps}.

The magnetic permeability tensor is calculated independently within a two-sublattice ferrimagnetic model based on the coupled Landau--Lifshitz--Gilbert equations, with contributions from the acoustic ferromagnetic resonance and the high-frequency Kaplan--Kittel exchange mode (Appendix~\ref{app:mu_tensor}). For BiYIG, the latter lies in the THz range but has an extremely small oscillator strength because of the near equality of the sublattice $g$-factors, $\left|g_1-g_2\right|\ll 1$ \cite{kaplan1953exchange,hsu2020observation}. The Kaplan–Kittel mode controls the resonant magnetic response in the THz range, whereas the broadband high-frequency tail of the permeability is predominantly determined by the much stronger acoustic mode. This treatment therefore provides a unified electrodynamic framework in which the electric-dipole and magnetic-dipole contributions to the magneto-optical response are obtained directly from $\varepsilon_{xy}(\omega)$ and $\mu_{xy}(\omega)$, respectively, rather than being introduced as independent phenomenological corrections \cite{pershan1967magnetooptical,krinchik1959tensors,gribova2024inverse,kang2012farinfrared}.

\begin{figure*}[htb!]
			\centering
			\begin{subfigure}[t]{0.47\textwidth}
				\centering
				\includegraphics[width=\textwidth]{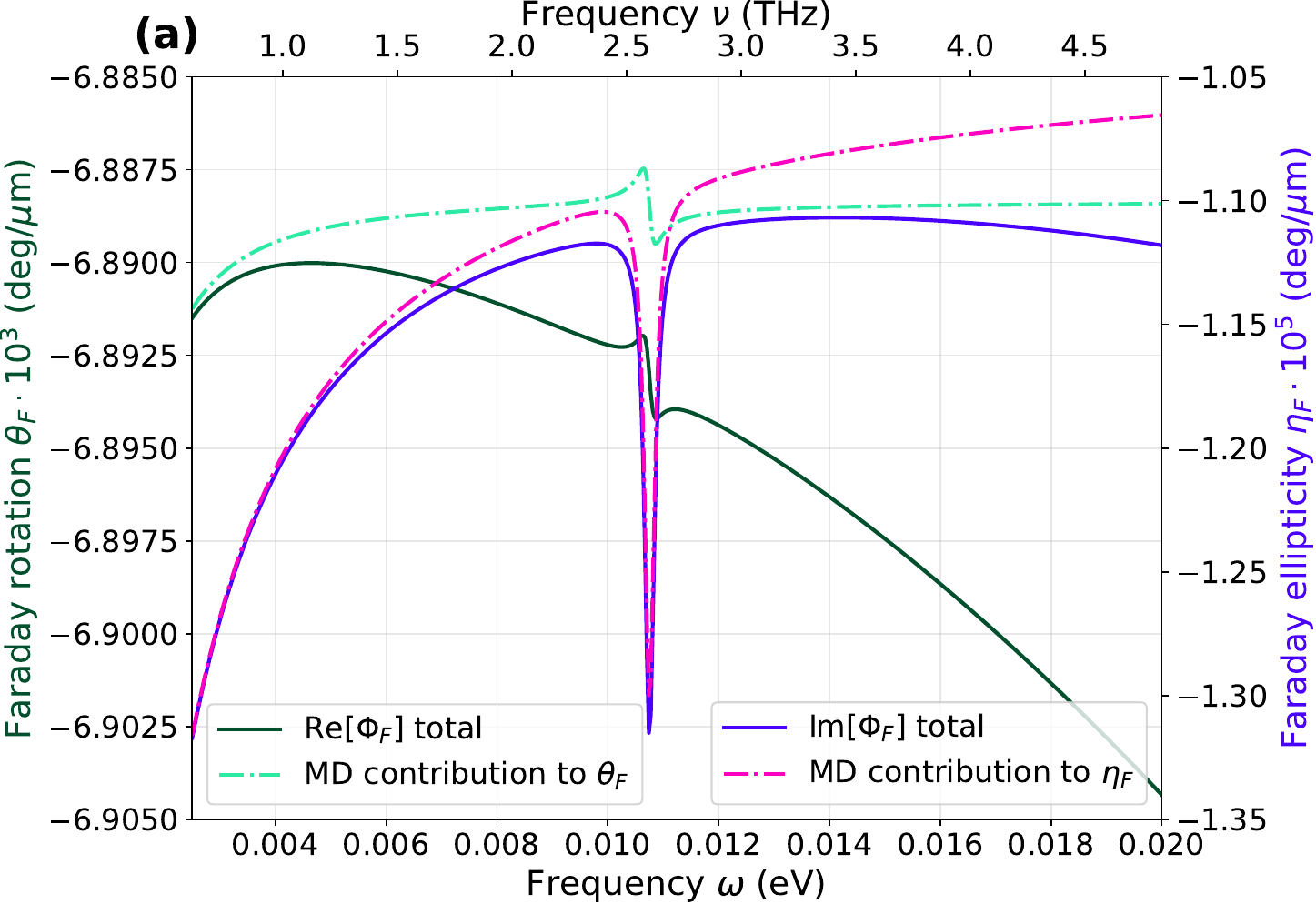}
				\label{fig:faraday_thz}
			\end{subfigure}
			\begin{subfigure}[t]{0.47\textwidth}
				\centering
				\includegraphics[width=\textwidth]{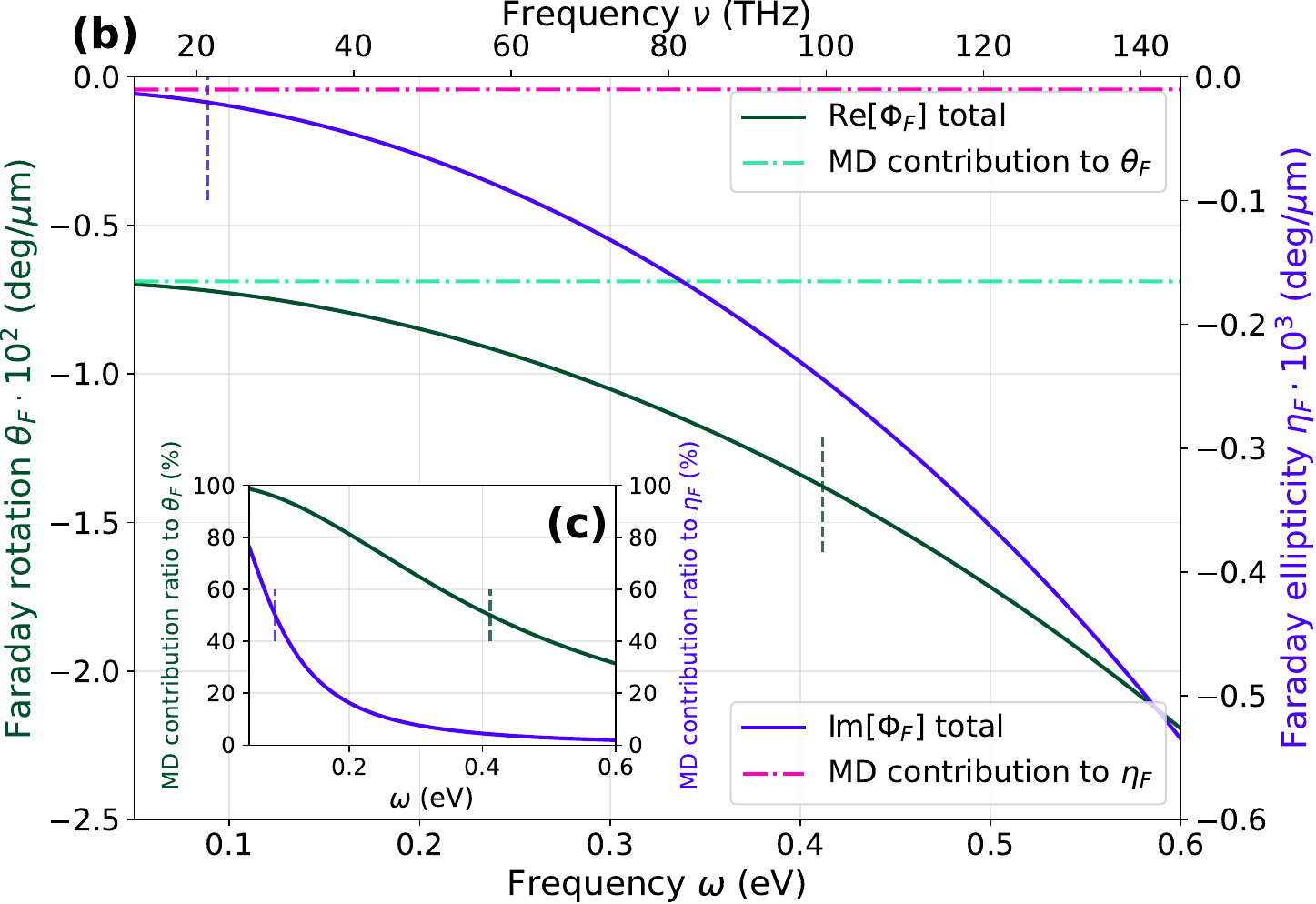}
				\label{fig:faraday_nir}
			\end{subfigure}
\caption{\justifying{\textbf{Spectral dispersion of the magneto-optical Faraday effect.} The analytical solutions within the first-order gyrotropic approximation (solid curves) are compared against the MD reference response (dash-dotted curves) for the Faraday rotation $\theta_F$ and ellipticity $\eta_F$. \textbf{(a)} Terahertz domain. The narrow resonance at $\omega \approx 0.0107$~eV (2.6~THz) corresponds to the direct excitation of the optical magnon mode (Kaplan-Kittel exchange resonance). \textbf{(b)} Near-infrared domain, demonstrating the crossover between MD and ED mechanisms. The vertical dashed lines indicate the critical frequencies where the MD contribution drops to 50\%: the crossover for ellipticity $\eta_F$ occurs at $\approx 0.09$~eV ($\approx 22$~THz), whereas for the rotation $\theta_F$, it occurs significantly later at $\approx 0.41$~eV ($\approx 99$~THz). \textbf{(c)} Inset quantitatively detailing the percentage ratio of the MD contribution to the total effect (0 to 100\%) across the MIR range.}}
\label{fig:faraday}
		\end{figure*}

Two spectral regions are considered in particular. The THz range is selected because the rapidly decreasing electric-dipole contribution is accompanied by the dispersive magnetic response associated with the Kaplan--Kittel and acoustic resonance, making this regime especially suitable for isolating the magnetic-dipole contribution. The MIR--NIR range is considered because it contains the crossover between the electric- and magnetic-dipole responses, allowing their relative magnitude to be identified explicitly; the importance of such a crossover has also been demonstrated experimentally in iron garnets in the NIR \cite{krinchik1960faraday,krinchik1961infrared,crossley1969faraday,gribova2024inverse}. The same frequency-dependent $\hat{\varepsilon}$ and $\hat{\mu}$ tensors are then used consistently to calculate the different magneto-optical geometries, so that the contribution of the Kaplan--Kittel mode is determined by a common material response rather than by geometry-specific fitting. In particular, the mode enters the magneto-optical observables through the off-diagonal permeability $\mu_{xy}$, while the conventional electric-dipole contribution is described by $\varepsilon_{xy}$. This tensorial separation makes it possible to identify directly the magnetic-dipole component of the Faraday, polar Kerr, transverse Kerr, longitudinal Kerr, and Cotton--Mouton effects, discussed in Sections~\ref{faraday_effect}, \ref{polar_kerr}, \ref{transverse_kerr}, \ref{longitudinal_kerr}, and \ref{cotton_mouton}, respectively.

Because the magneto-optical observables are nonlinear functions of the constitutive tensors, an additive separation into ED and MD contributions is not unique for all geometries. The comparison used below is therefore defined operationally. For any magneto-optical observable $\mathcal O$, the full response is calculated using the complete frequency-dependent tensors,
\[
\mathcal O_{\mathrm{full}}
=
\mathcal F[
\varepsilon_{xx}(\omega),
\varepsilon_{xy}(\omega),
\mu_{xx}(\omega),
\mu_{xy}(\omega)
].
\]
The MD reference response is obtained by suppressing the gyroelectric contribution and freezing the diagonal dielectric response at its zero-frequency model value,
\[
\mathcal O_{\mathrm{MD}}^{\mathrm{ref}}
=
\mathcal F[
\varepsilon_{xx}(0),
0,
\mu_{xx}(\omega),
\mu_{xy}(\omega)
].
\]
This construction removes the dispersive electronic contribution while retaining the full frequency-dependent magnetic response. 


\subsection{Magnetic circular birefringence: the Faraday effect}\label{faraday_effect}

The Faraday effect provides a direct probe of the circular birefringence and dichroism of a magnetized medium and is well established in the visible and near-infrared spectral ranges \cite{gribova2024inverse,krinchik1960faraday,vojna2020faraday}. At lower photon energies, however, its conventional description in terms of the dielectric tensor alone becomes incomplete because the magnetic response of the medium can no longer be neglected \cite{krinchik1959tensors,gribova2024inverse}. Here, the Faraday effect is calculated using the frequency-dependent dielectric and magnetic-permeability tensors introduced above, allowing the electric-dipole (ED) and magnetic-dipole (MD) contributions to be separated explicitly. The latter is determined by the off-diagonal permeability $\mu_{xy}$ and therefore contains the dynamics of both ferrimagnetic modes. In particular, the high-frequency Kaplan--Kittel mode produces a narrow resonance in the THz range, whereas its small oscillator strength strongly limits the associated absorption \cite{kaplan1953exchange,hsu2020observation}. This makes the Faraday effect especially suitable for identifying the MD contribution away from the narrow resonance itself and for tracking its crossover with the ED contribution toward the NIR.

Let's consider a magnetized medium with $\mathbf{M}\parallel z$ and light incident in the $xz$ plane. The complex refractive indices of the two counter-rotating circular eigenmodes are determined from Maxwell's equations using the constitutive tensors $\hat{\varepsilon}$ and $\hat{\mu}$. For a plane wave
$\mathbf{E}(\mathbf{r},t)=\mathbf{E}_0\exp[i(\omega\mathbf{n}\cdot\mathbf{r}/c-\omega t)]$,
the field relations $\hat{\mu}\mathbf{H}=\mathbf{n}_t^\pm\times\mathbf{E}$ and $\hat{\varepsilon}\mathbf{E}=-\mathbf{n}_t^\pm\times\mathbf{H}$ yield the exact dispersion relation for
$\mathbf{n}_t^\pm=(n_{t,x},0,n_{t,z}^{\pm})$:
\begin{equation}
	\varepsilon_{xx}\mu_{xx} \Bigl(1-\frac{(\epsilon+\epsilon_\mu)^2}{4}\Bigr)-n_{t,x}^2
	=
	\Bigl(n_{t,z}^{\pm}\mp i\frac{\sqrt{\varepsilon_{xx}\mu_{xx}}}{2}(\epsilon+\epsilon_\mu)\Bigr)^2,
	\label{n_z:faraday}
\end{equation}
where $\epsilon = {\varepsilon_{xy}}/{\varepsilon_{xx}}$, $\epsilon_\mu= {\mu_{xy}}/{\mu_{xx}}$, and $n_{t,x}$ is conserved across the interface by phase matching.

For normal incidence, $n_{t,x}=0$, the complex specific Faraday angle is \cite{zvezdin1997modern}
\begin{equation}
	\Phi_F = \theta_F + i \eta_F =  -\frac{\pi }{\lambda} (n_{t,z}^+ - n_{t,z}^-),
\end{equation}
where $\theta_F$ is the specific Faraday rotation, and $\eta_F$ is the specific ellipticity arising from circular dichroism.

In the weak-gyrotropy limit, when
$|\varepsilon_{xx}^{\prime}|\gg \max\{|\varepsilon_{xx}^{\prime\prime}|,|\varepsilon_{xy}^{\prime}|,|\varepsilon_{xy}^{\prime\prime}|\}$
and
$|\mu_{xx}^{\prime}|\gg \max\{|\mu_{xx}^{\prime\prime}|,|\mu_{xy}^{\prime}|,|\mu_{xy}^{\prime\prime}|\}$,
the exact expression reduces to
\begin{equation}
	\Phi_F \approx -i \frac{\pi l\sqrt{\varepsilon_{xx}\mu_{xx}}}{\lambda}\biggl(\frac{\varepsilon_{xy}}{\varepsilon_{xx}} + \frac{\mu_{xy}}{\mu_{xx}}\biggr).\label{complex_Faraday}
\end{equation}
Equation~\ref{complex_Faraday} provides the central decomposition used below: the first term represents the ED contribution, while the second originates from the MD response. This form is consistent with the generalization of the Faraday effect to media with both gyroelectric and gyromagnetic response \cite{krinchik1959tensors,gribova2024inverse,zvezdin1997modern}. The relevance of the permeability contribution at long wavelengths has been established experimentally in iron garnets \cite{krinchik1961infrared,crossley1969faraday} and in the NIR analysis of iron garnet films \cite{gribova2024inverse}.

To isolate the MD contribution, the static diagonal permittivity $\varepsilon_{xx}(0)$ is obtained from the multi-resonance description of the experimental optical spectra rather than from an independent background estimate. The fitted resonances reproduce the measured visible spectrum, while their low-energy tails provide the extrapolation required in the NIR and THz ranges (Appendix~\ref{tensor_eps}). This construction preserves the dispersive ED response in the full calculation. In contrast, the MD reference response is obtained by setting $\varepsilon_{xy}=0$ and replacing $\varepsilon_{xx}(\omega)$ by $\varepsilon_{xx}(0)$ in Eq.~\ref{complex_Faraday}. Figure~\ref{fig:faraday} therefore directly compares the complete response with the contribution arising from $\mu_{xy}$ alone.

In the THz range [Fig.~\ref{fig:faraday}(a)], the two-sublattice permeability produces a pronounced dispersive feature at $\hbar\omega\approx0.0107$~eV ($\nu\approx2.6$~THz), corresponding to the Kaplan--Kittel exchange mode of BiYIG. The mode results from the out-of-phase dynamics of the two Fe$^{3+}$ sublattices and is described by the second term in the permeability tensor \cite{kaplan1953exchange,hsu2020observation}. Its oscillator strength is strongly suppressed by the near equality of the sublattice $g$-factors, as quantified by the parameters of the two-sublattice model (Appendix~\ref{app:mu_tensor}). Consequently, the resonance produces only a narrow absorptive feature, whereas outside this narrow spectral interval the permeability response is predominantly dispersive. In this sub-resonant THz regime, the ED contribution is strongly suppressed and the Faraday response is therefore governed by the MD term $\mu_{xy}/\mu_{xx}$.  This is consistent with the established importance of magnetic-permeability effects in the infrared and with the observed connection between Kaplan--Kittel modes and magnetic permeability in ferrimagnetic garnets \cite{krinchik1962exchange,kang2012farinfrared,gribova2024inverse}.

The NIR range [Fig.~\ref{fig:faraday}(b)] provides a different regime in which the MD and ED contributions become comparable. The MD reference response varies slowly with frequency, whereas the ED contribution retains the dispersive tail of the electronic resonances included in $\varepsilon_{xy}(\omega)$. Their relative weight therefore changes continuously with photon energy. For the Faraday rotation, the two contributions become equal at $\hbar\omega\approx0.41$~eV ($\nu\approx99$~THz), where the MD contribution accounts for 50\% of the total rotation. The corresponding crossover in the ellipticity occurs at the substantially lower frequency $\hbar\omega\approx0.09$~eV ($\nu\approx22$~THz). This difference follows from the distinct sensitivity of rotation and ellipticity to the dispersive and absorptive parts of the constitutive tensors. Thus, the NIR regime provides a direct measure of the crossover between the ED and MD mechanisms, whereas the THz regime isolates the MD-dominated response.

\begin{figure*}[htb!]
    \centering
\begin{subfigure}[t]{0.47\textwidth}
				\centering
				\includegraphics[width=\textwidth]{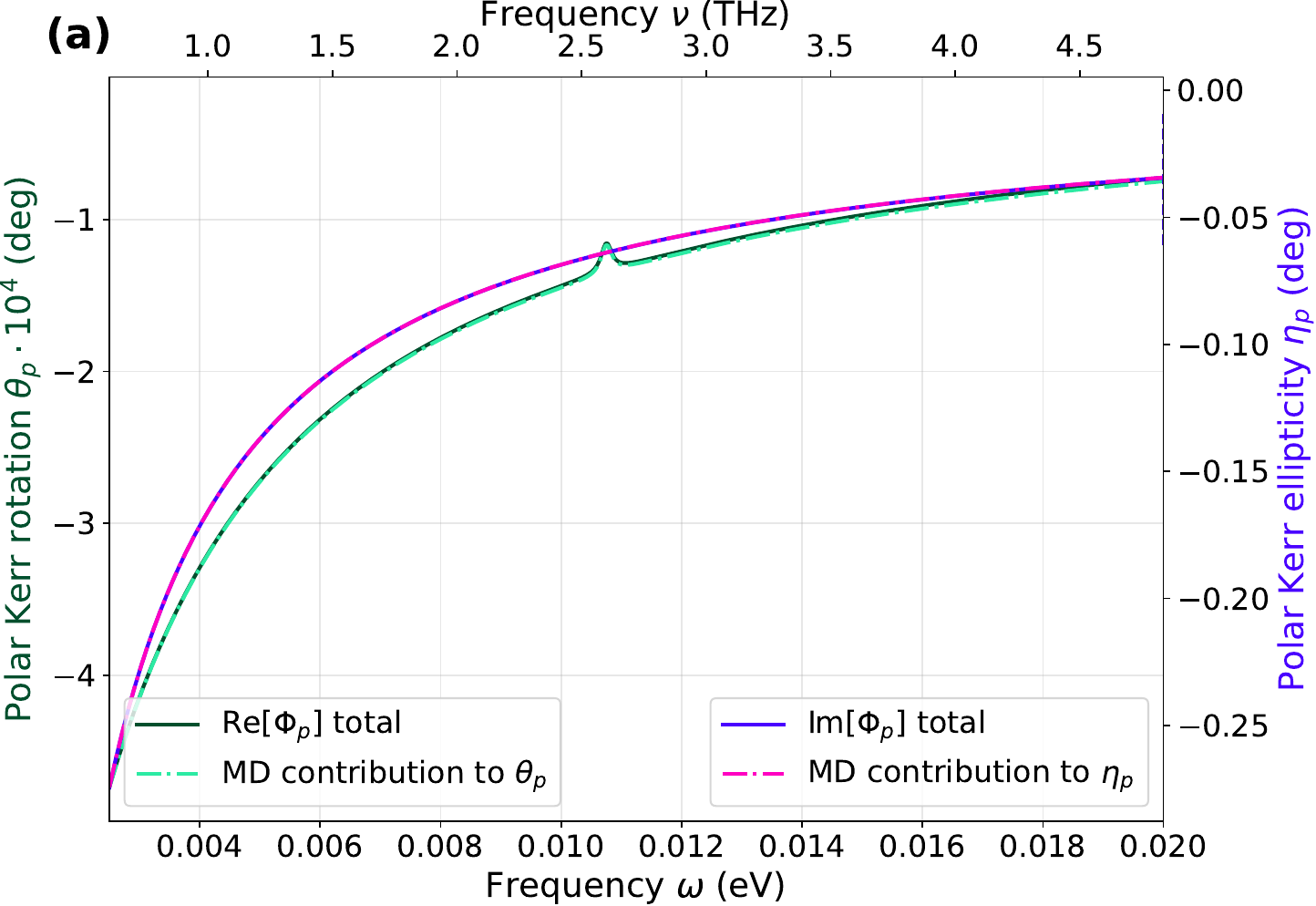}
				\label{fig:kerr_thz}
			\end{subfigure}
			\begin{subfigure}[t]{0.47\textwidth}
				\centering
				\includegraphics[width=\textwidth]{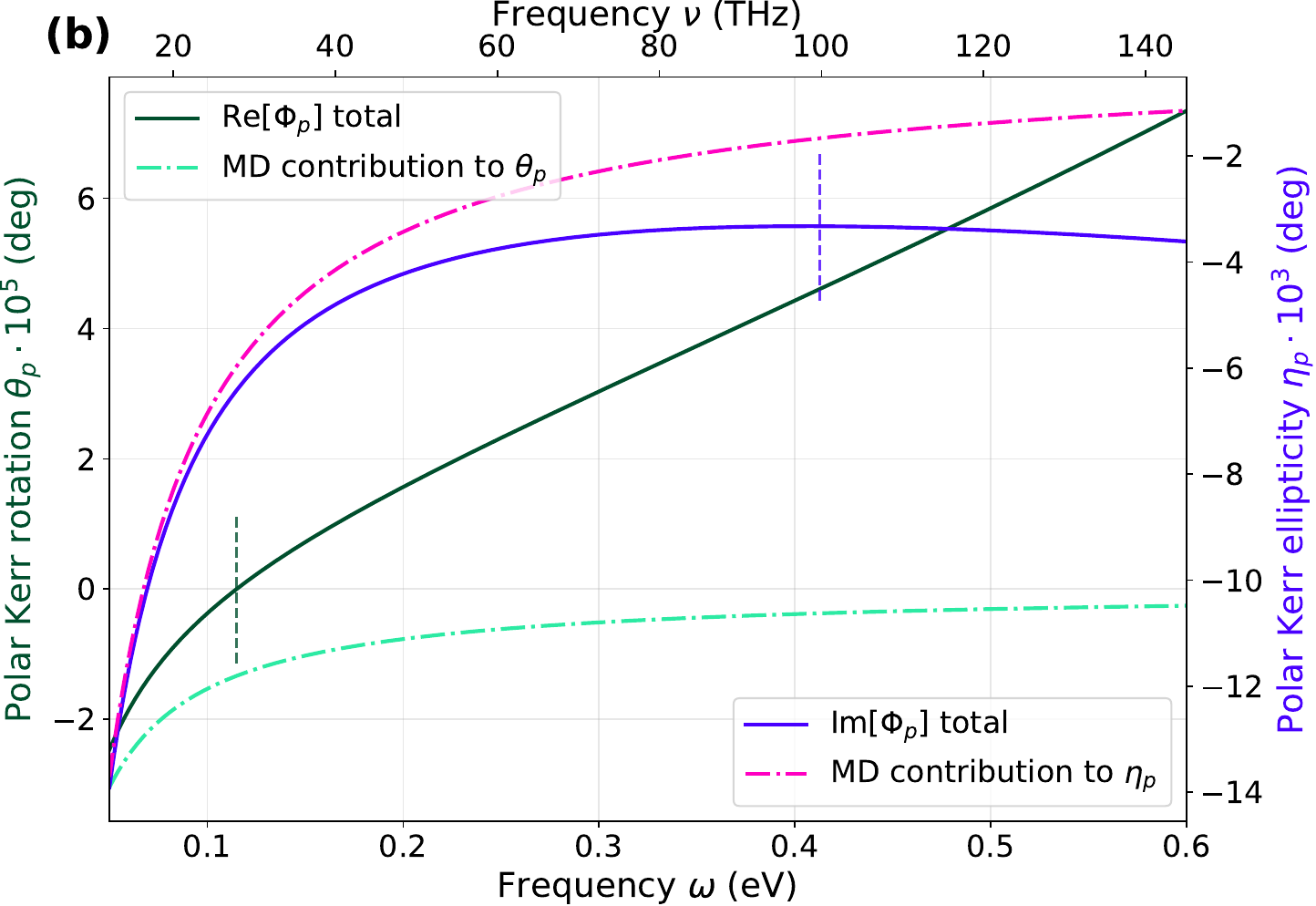}
				\label{fig:kerr_nir}
			\end{subfigure}    \caption{\justifying{\textbf{Spectral dispersion of the polar Kerr effect}, comparing the analytical solution within the first-order gyrotropic approximation (solid curves) with the MD reference response (dash-dotted curves) for Kerr rotation $\theta_p$ and ellipticity $\eta_p$. \textbf{(a)} Terahertz domain. The optical magnon mode (exchange resonance) at $0.0107$~eV ($2.6$~THz) manifests exclusively as a sharp peak in the rotation $\theta_p$, reflecting the $90^\circ$ phase-shifted boundary conditions of the reflection geometry compared to transmission. \textbf{(b)} Near-infrared domain showing the magnitude crossover between MD and ED mechanisms. Vertical dashed lines indicate the frequencie, where the MD-reference response reaches 50\% of the full response: $\approx 0.11$~eV for $\theta_p$ and $\approx 0.41$~eV for $\eta_p$. This demonstrates an inverted spectral hierarchy of the magnetic response compared to the Faraday effect.}}
    \label{fig:polar_kerr}
\end{figure*}
\subsection{Polar Kerr effect}\label{polar_kerr}

The Faraday effect probes magneto-optical propagation through the bulk of the sample, whereas the polar Kerr effect probes the same material response through reflection at the sample--vacuum interface. The two geometries therefore provide complementary access to the same electric- and magnetic-dipole mechanisms, while the boundary conditions can change how their dispersive and absorptive components appear in the measured rotation and ellipticity.

The normal incidence on the sample surface is considered with the magnetization oriented along the surface normal, corresponding to the polar Kerr geometry. For light propagating from vacuum into the magnetized medium, the complex amplitude reflection coefficients for the right- and left-circularly polarized eigenmodes are
\begin{equation}
	r_{\pm}=\frac{n_{t,z}^{\pm}-1}{n_{t,z}^{\pm}+1}.
\end{equation}

The complex polar Kerr angle, $\Phi_p=\theta_p+i\eta_p$, is determined by the difference between the two circular reflection coefficients \cite{zvezdin1997modern}:
\begin{equation}
	\Phi_p = \theta_p + i\eta_p= -i\frac{r_+ - r_-}{r_+ + r_-}\approx -\frac{\varepsilon_{xy}\mu_{xx}+\varepsilon_{xx}\mu_{xy}}{\sqrt{\varepsilon_{xx}\mu_{xx}}(1-\varepsilon_{xx}\mu_{xx})}.\label{eq:polar}
\end{equation}
This expression contains the ED contribution through $\varepsilon_{xy}$ and the MD contribution through $\mu_{xy}$. In contrast to the Faraday geometry, these contributions enter the reflection coefficient through the interface boundary conditions rather than through phase accumulation over the propagation length. Consequently, the relative phase of the dispersive and absorptive contributions to $\theta_p$ and $\eta_p$ differs from that in the Faraday effect.

The off-diagonal magnetic permeability has previously been shown to produce a Kerr response in engineered magneto-plasmonic metamaterials \cite{yang2022gyromagnetic}. Here, the corresponding intrinsic gyromagnetic response of a ferrimagnetic garnet and its evolution is investigated from the THz to the NIR spectral range.

In the THz range [Fig.~\ref{fig:polar_kerr}(a)], our calculations show that the polar Kerr response is dominated by the MD contribution throughout the spectral region outside the narrow Kaplan--Kittel resonance. The two-sublattice permeability produces the same exchange-mode resonance at $\hbar\omega\approx0.0107$~eV ($\nu\approx2.6$~THz) identified in the Faraday response. Because the Kaplan--Kittel mode has a strongly suppressed oscillator strength, the associated absorption remains confined to a narrow spectral interval, while the surrounding response is predominantly dispersive \cite{kaplan1953exchange,hsu2020observation}. In the polar Kerr geometry, however, the resonance appears predominantly in the Kerr rotation $\theta_p$, whereas the ellipticity $\eta_p$ remains smooth through the resonance. This redistribution of the resonant feature between rotation and ellipticity follows directly from the complex reflection coefficient in Eq.~\ref{eq:polar}: the interface response introduces a different phase relation between the real and imaginary parts of $\mu_{xy}$ than in the transmitted Faraday signal. Thus, the same underlying magnetic resonance produces qualitatively different observables in transmission and reflection, without changing the underlying material response.

In the MIR--NIR range [Fig.~\ref{fig:polar_kerr}(b)], the MD contribution again competes with the dispersive ED background, but the crossover occurs at different energies for $\theta_p$ and $\eta_p$. An equal-magnitude crossover between the MD-reference and ED-associated to the Kerr rotation become equal in magnitude at $\hbar\omega\approx0.11$~eV ($\nu\approx27$~THz), whereas the corresponding crossover in the Kerr ellipticity occurs at $\hbar\omega\approx0.41$~eV ($\nu\approx99$~THz). The difference from the Faraday response follows from the reflection geometry rather than from a change in the material tensors: the same $\varepsilon_{xy}(\omega)$ and $\mu_{xy}(\omega)$ enter Eq.~\ref{eq:polar}, but their contributions are weighted differently by the interface response. The persistence of a substantial MD contribution into the NIR is consistent with earlier infrared studies of iron garnets and with recent analyses showing that the permeability contribution must be retained in the NIR magneto-optical response \cite{krinchik1961infrared,crossley1969faraday,gribova2024inverse}. The different crossover frequencies obtained for $\theta_p$ and $\eta_p$ therefore demonstrate that the apparent disappearance of the MD response cannot be characterized by a single spectral cutoff; it depends on the magneto-optical observable and on the measurement geometry.

\begin{figure*}[htb!]
\centering
\begin{subfigure}[t]{0.47\textwidth}
				\centering
				\includegraphics[width=\textwidth]{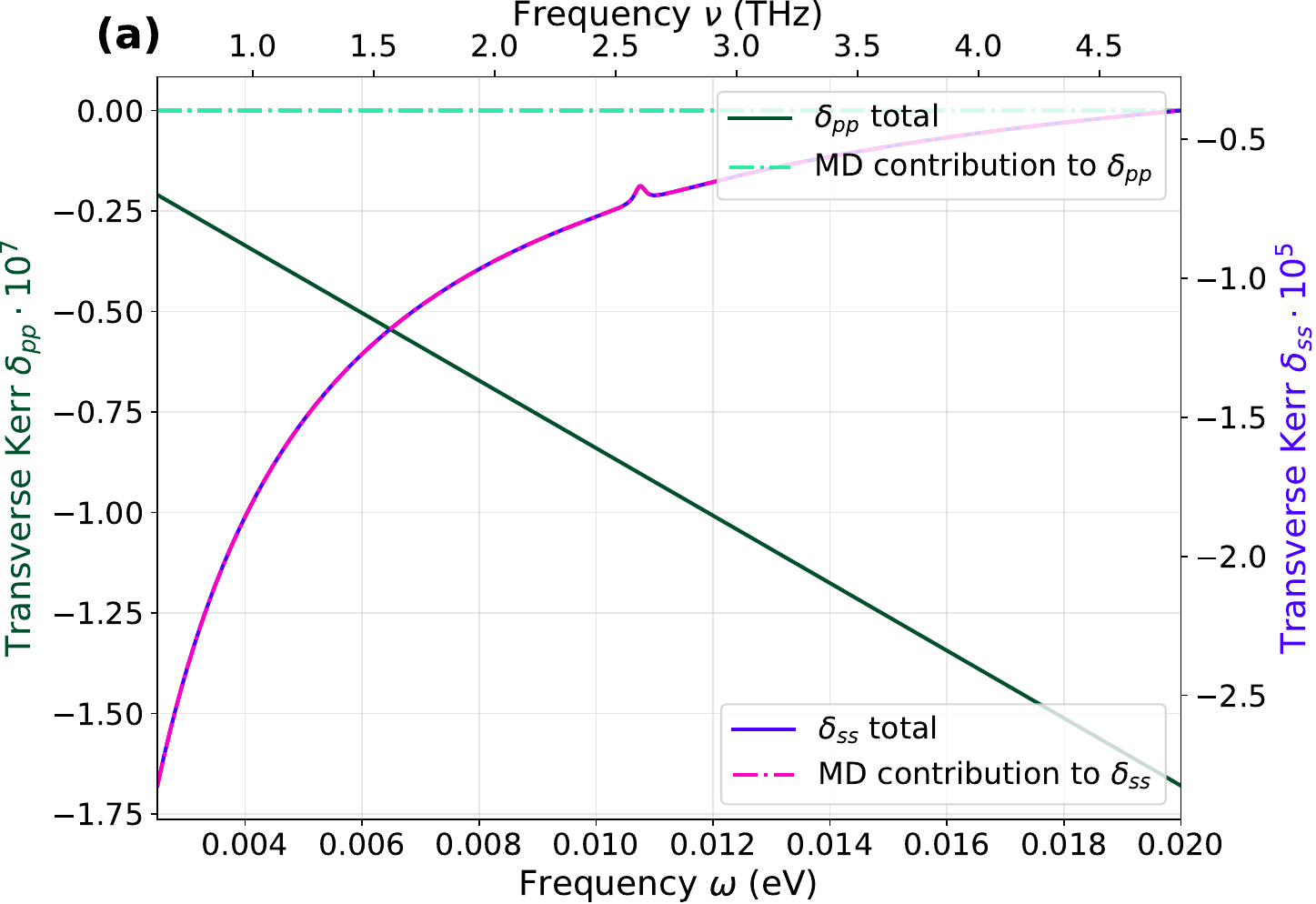}
				\label{fig:kerr_thz}
			\end{subfigure}
			\begin{subfigure}[t]{0.47\textwidth}
				\centering
				\includegraphics[width=\textwidth]{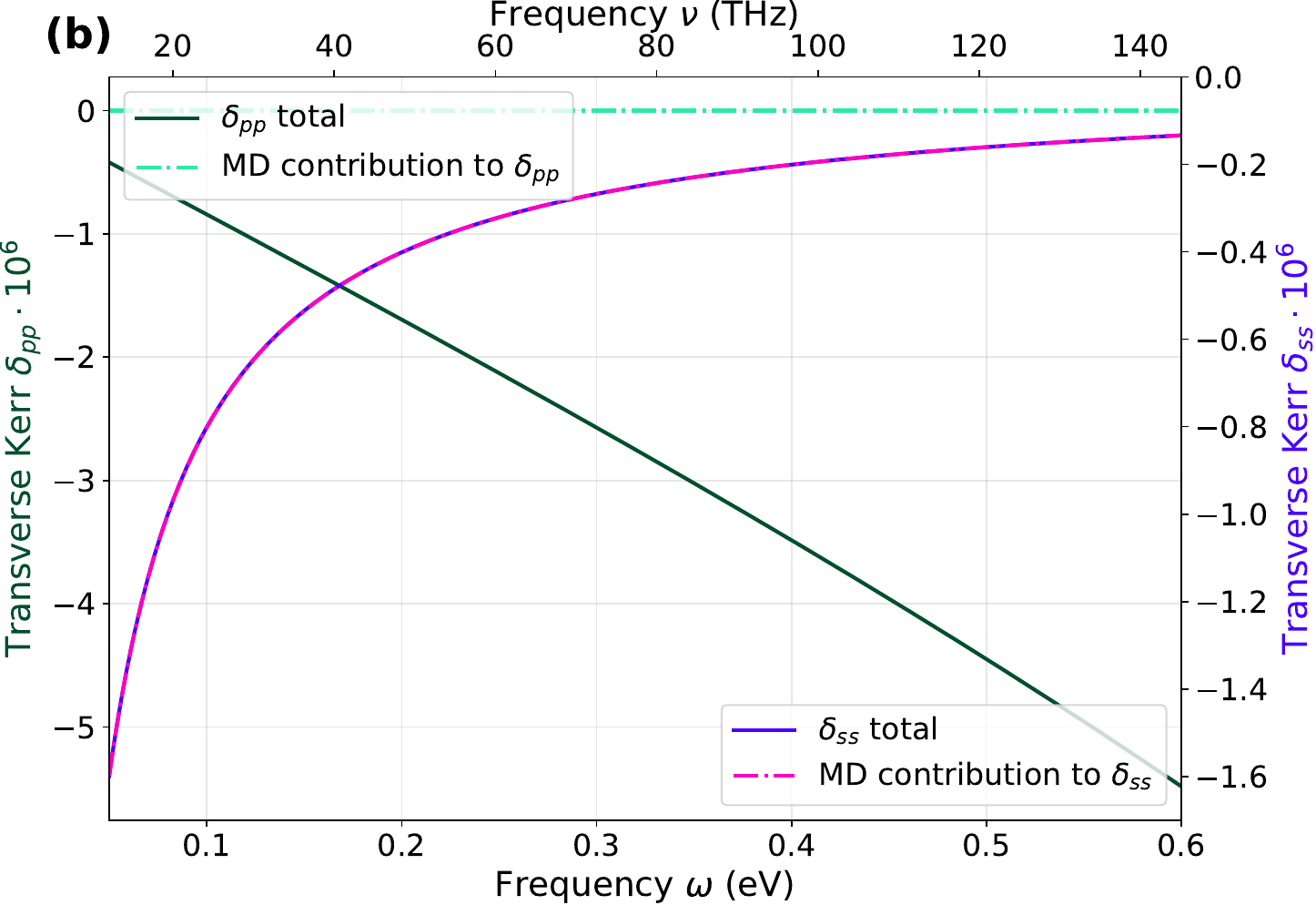}
				\label{fig:kerr_nir}
			\end{subfigure}  
\caption{\justifying{\textbf{Spectral dispersion of the Transverse Kerr effect} for p-polarized ($\delta_{pp}$, dark green) and s-polarized ($\delta_{ss}$, purple) incident light. The solid curves represent the total analytical solutions, while the dash-dotted lines denote the MD reference response. \textbf{(a)} Terahertz domain. The magnitude of $\delta_{ss}$ is two orders of magnitude larger than $\delta_{pp}$. The Kaplan-Kittel optical magnon mode at $\approx 0.0107$~eV ($2.6$~THz) is exclusively resolved in the s-polarization, confirming $\delta_{ss}$ as a pure, background-free probe for spin dynamics. \textbf{(b)} Near-infrared domain. While both effects are of the same order of magnitude ($\sim 10^{-6}$), their absolute values exhibit opposite trends: the MD-governed $\delta_{ss}$ decays as the frequency moves away from the magnetic resonances, whereas the ED-governed $\delta_{pp}$ grows as it approaches the electronic transitions.}}
\label{fig:tke}
\end{figure*}

\subsection{Transverse Kerr effect}\label{transverse_kerr}

The Faraday and polar Kerr results show that the magnetic-permeability contribution remains relevant in the THz--NIR range. The transverse Kerr geometry provides a more selective test of this contribution because, for $s$-polarized incidence, the magnetization-dependent reflection is coupled to the off-diagonal permeability rather than to the off-diagonal permittivity. This gyromagnetic selection rule was demonstrated experimentally in the NIR using a magneto-plasmonic metamaterial \cite{yang2022gyromagnetic}. The present calculation addresses a different physical situation: the gyromagnetic response is generated intrinsically by the two-sublattice ferrimagnetic dynamics of BiYIG, rather than by an engineered metamaterial resonance, and is followed from the THz to the NIR spectral range.

Consider $p$-polarized (TM, Fig.~\ref{fig:middle}) and $s$-polarized (TE) waves incident on the sample surface. The plane of incidence is the $xOy$ plane, and $\varphi$ denotes the angle of incidence measured from the surface normal. The transmitted wave propagates inside the medium at an angle $\varphi_t$. In this geometry the polarization states remain decoupled. The refractive indices of the transmitted $p$- and $s$-polarized waves are
\begin{align}
n_t^p &= \sqrt{\varepsilon_{xx}\mu_{zz}}\Bigl(1+\frac{\epsilon^2}{2}\Bigr), \label{eq:nt_p}\\
n_t^s &= \sqrt{\varepsilon_{zz}\mu_{xx}}\Bigl(1+\frac{\epsilon^2_\mu}{2}\Bigr). \label{eq:nt_s}
\end{align}

The reflection coefficient for incident $p$-polarized light is
\begin{equation}
r_{pp} = \frac{\cos \varphi (\cos\varphi_t-\epsilon\sin\varphi_t)-\sqrt{\mu_{zz}/\varepsilon_{xx}}\cos^2\varphi_t}{\cos \varphi (\cos\varphi_t-\epsilon\sin\varphi_t)+\sqrt{\mu_{zz}/\varepsilon_{xx}}\cos^2\varphi_t}.
\label{eq:transverse_rpp}
\end{equation}

For incident $s$-polarized light,
\begin{equation}
r_{ss} = \frac{\cos \varphi (\cos\varphi_t-\epsilon_\mu\sin\varphi_t)-\sqrt{\varepsilon_{zz}/\mu_{xx}}\cos^2\varphi_t}{\cos \varphi (\cos\varphi_t-\epsilon_\mu\sin\varphi_t)+\sqrt{\varepsilon_{zz}/\mu_{xx}}\cos^2\varphi_t}.
\label{eq:transverse_rss}
\end{equation}
Here,
\[
\epsilon=\frac{\varepsilon_{xy}}{\varepsilon_{xx}},
\qquad
\epsilon_\mu=\frac{\mu_{xy}}{\mu_{xx}}.
\]

The magnetization dependence follows from the Onsager symmetry of the constitutive tensors: the diagonal components are even in $\mathbf M$, whereas the off-diagonal components are odd. Therefore,
\[
\epsilon(-\mathbf M)=-\epsilon(\mathbf M),
\qquad
\epsilon_\mu(-\mathbf M)=-\epsilon_\mu(\mathbf M),
\]
while the diagonal elements remain unchanged under magnetization reversal. The terms linear in $\epsilon$ and $\epsilon_\mu$ in Eqs.~(\ref{eq:transverse_rpp}) and (\ref{eq:transverse_rss}) consequently produce different reflection intensities for opposite magnetization directions. The transverse Kerr signals are defined as
\begin{align}
\delta_{pp} &= \frac{|r_{pp}(\mathbf{M})|^2 - |r_{pp}(-\mathbf{M})|^2}{|r_{pp}(0)|^2}, \label{eq:delta_pp} \\
\delta_{ss} &= \frac{|r_{ss}(\mathbf{M})|^2 - |r_{ss}(-\mathbf{M})|^2}{|r_{ss}(0)|^2}. \label{eq:delta_ss}
\end{align}

The essential difference between the two polarization channels is already contained in Eqs.~(\ref{eq:transverse_rpp}) and (\ref{eq:transverse_rss}). The magnetization-dependent part of $r_{pp}$ contains the gyroelectric parameter $\epsilon$, whereas $r_{ss}$ contains $\epsilon_\mu$ instead of $\epsilon$. Thus,
\begin{equation}
\mu_{xy}=0 \quad \Longrightarrow \quad \delta_{ss}=0,
\end{equation}
irrespective of $\varepsilon_{xy}$. The $s$-polarized TMOKE therefore excludes the off-diagonal electric-dipole contribution and selectively probes the gyromagnetic response through $\mu_{xy}$. The ordinary diagonal permittivity and permeability still determine the Fresnel background, but the magnetization-dependent $s$-polarized term is controlled by the off-diagonal permeability alone. This selection rule is consistent with the experimental demonstration of $s$-polarized TMOKE in a gyromagnetic magneto-plasmonic metamaterial by Yang \textit{et al.} \cite{yang2022gyromagnetic}.

Figure~\ref{fig:tke} shows the calculated $\delta_{pp}$ and $\delta_{ss}$ at $\varphi=45^\circ$. In the THz range [Fig.~\ref{fig:tke}(a)], the two polarization channels separate by their physical origin. The $s$-polarized response is governed entirely by $\mu_{xy}$ and reaches approximately $10^{-5}$, whereas the $p$-polarized response, which contains the ED contribution, is approximately $10^{-7}$. Thus, within the present model, the $s$-polarized TMOKE is about two orders of magnitude larger than the $p$-polarized signal in the THz range. More importantly, only $\delta_{ss}$ exhibits the narrow anomaly at $\hbar\omega\approx0.0107$~eV ($\nu\approx2.6$~THz), corresponding to the Kaplan--Kittel exchange mode. The same mode is obtained from the two-sublattice permeability model used in the Faraday and polar-Kerr calculations and is strongly suppressed in oscillator strength by the near equality of the sublattice $g$-factors. The resonance therefore appears as a narrow feature in the $s$-polarized response, while $\delta_{pp}$ remains smooth. The THz result demonstrates that the transverse Kerr geometry can separate the intrinsic magnetic-dipole response from the electric-dipole background and provide direct optical access to the Kaplan--Kittel contribution.

The relation to the work of Yang \textit{et al.} is therefore specific. Their experiment demonstrated the same $s$-polarized TMOKE selection rule in the NIR and attributed the observed signal to a nonzero off-diagonal permeability in a magneto-plasmonic metamaterial \cite{yang2022gyromagnetic}. The present result does not introduce a different selection rule and does not contradict that observation. Instead, it shows that the same electrodynamic channel arises intrinsically in a ferrimagnetic garnet, where $\mu_{xy}(\omega)$ is determined by the coupled dynamics of the tetrahedral and octahedral Fe$^{3+}$ sublattices. In contrast to the engineered gyromagnetic response of the metamaterial, the frequency dependence here is therefore fixed by the intrinsic ferrimagnetic modes, including the Kaplan--Kittel exchange resonance. The calculated THz signal extends the role of the $s$-polarized TMOKE from the previously demonstrated NIR gyromagnetic response to the intrinsic high-frequency magnetic dynamics of BiYIG. The findings of this work also show that in Thz region the signal from $\delta_{ss}$ is in two orders higher than the $\delta_{pp}$.

The NIR response [Fig.~\ref{fig:tke}(b)] provides the complementary regime in which the two channels become comparable in magnitude. The $s$-polarized signal remains purely MD in origin and decreases with increasing photon energy as the frequency moves away from the low-energy magnetic resonances. In contrast, the $p$-polarized response increases toward the electronic resonances because it contains the dispersive off-diagonal permittivity $\varepsilon_{xy}$. The two signals reach comparable magnitudes of order $10^{-6}$ in the MIR/NIR, but their opposite spectral trends distinguish their origins. This behavior is consistent with the previously established importance of magnetic-permeability contributions to the infrared magneto-optical response of ferrite garnets \cite{krinchik1961infrared,crossley1969faraday,gribova2024inverse}. The polarization dependence therefore provides an independent separation of the two mechanisms: $\delta_{ss}$ isolates the gyromagnetic contribution, whereas $\delta_{pp}$ retains the competition between gyroelectric and gyromagnetic responses.

The resulting comparison with \cite{yang2022gyromagnetic} establishes a direct connection between polarization-selective TMOKE and the tensorial origin of the magneto-optical response. The $s$-polarized channel isolates $\mu_{xy}$, making it particularly sensitive to the ferrimagnetic dynamics in the THz range, while the $p$-polarized channel probes the conventional electric-dipole response and its interference with the magnetic contribution. The novelty is not the gyromagnetic selection rule itself, but its application to the intrinsic two-sublattice permeability of BiYIG, revealing an \(s\)-polarized TMOKE channel that directly tracks the intrinsic ferrimagnetic dynamics and, in particular, the Kaplan--Kittel exchange mode in the THz range.

\begin{figure*}[htb!]
    \centering
\begin{subfigure}[t]{0.47\textwidth}
				\centering
				\includegraphics[width=\textwidth]{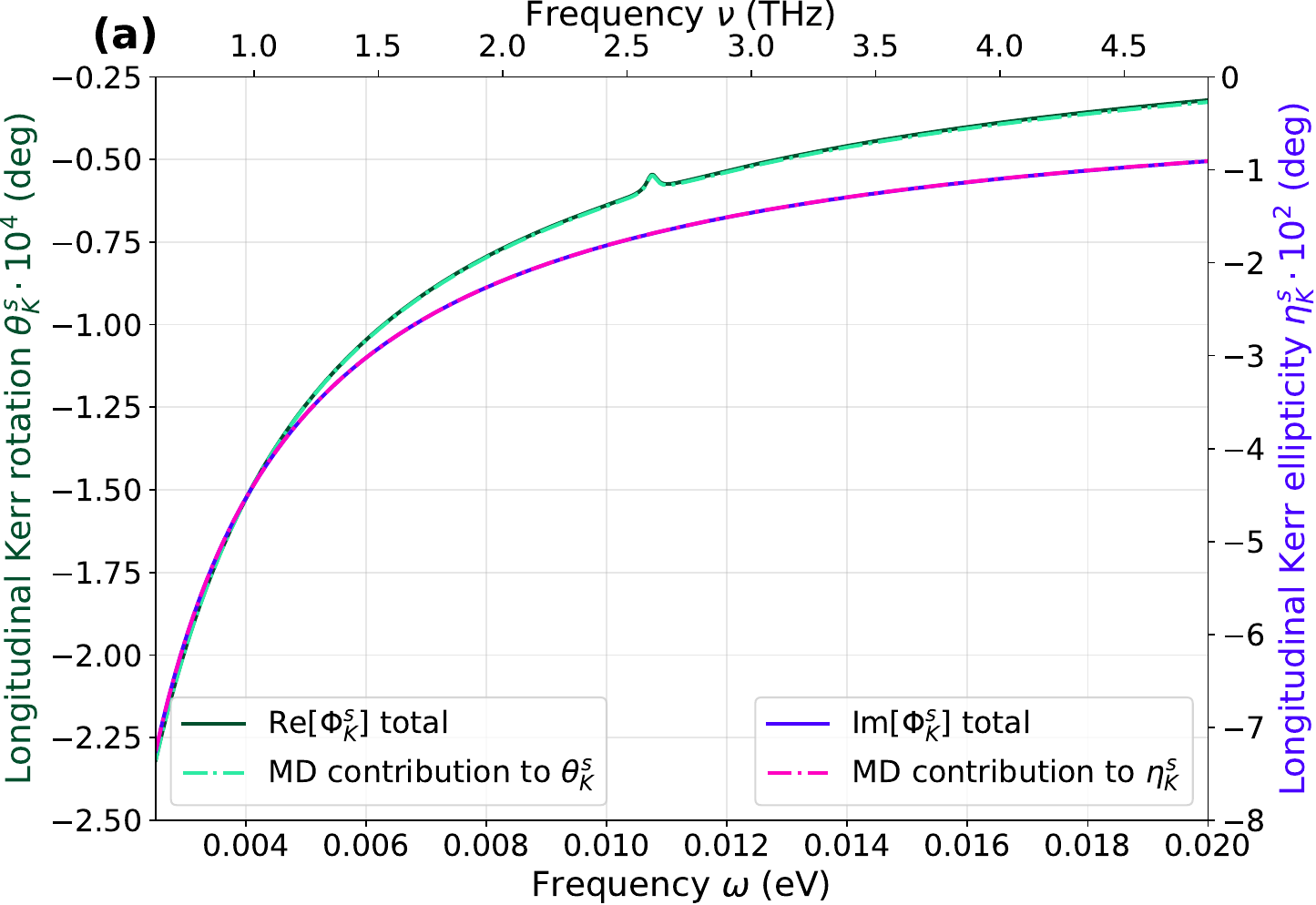}
				\label{fig:l_kerr_thz}
			\end{subfigure}
			\begin{subfigure}[t]{0.47\textwidth}
				\centering
				\includegraphics[width=\textwidth]{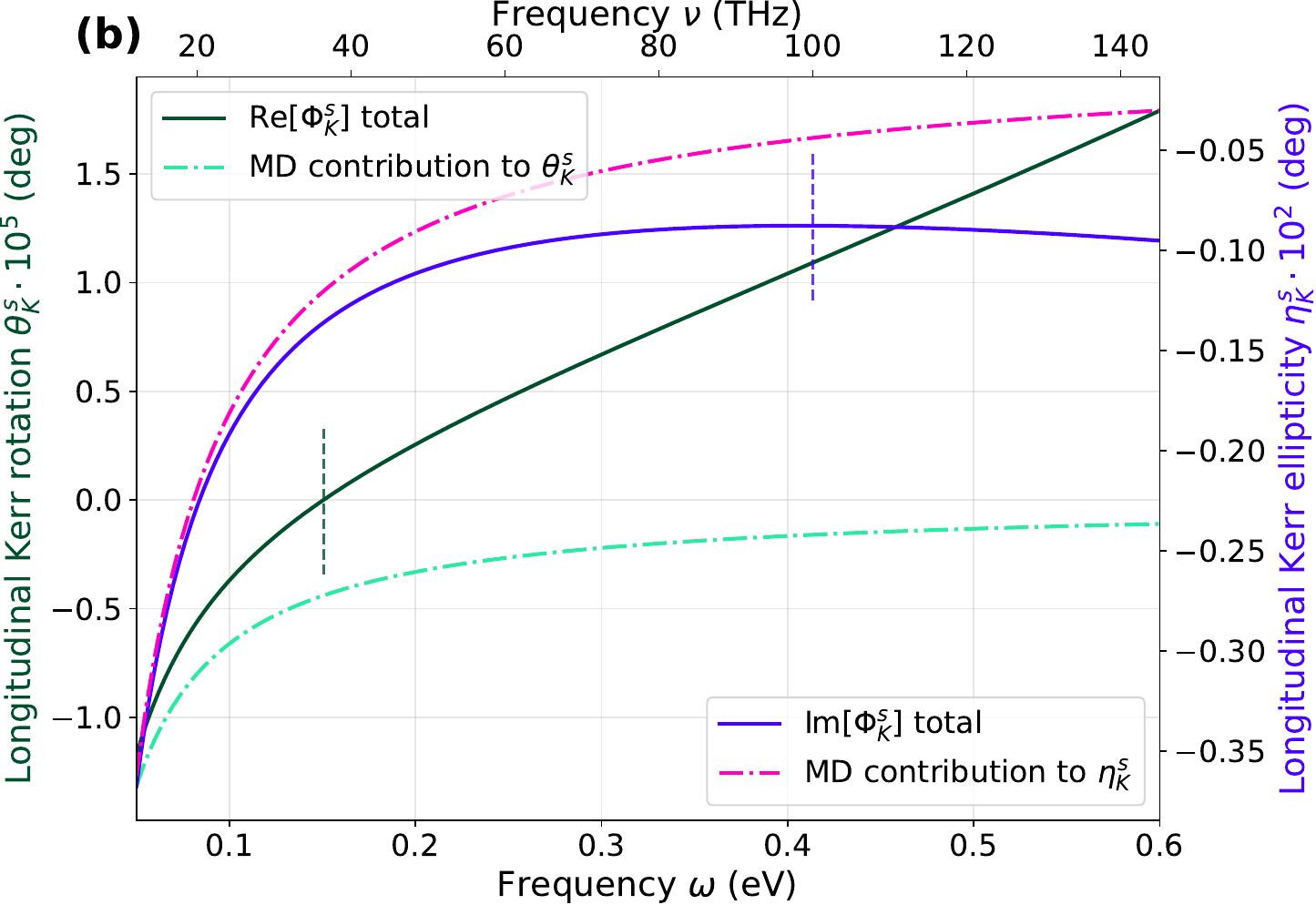}
				\label{fig:l_kerr_nir}
			\end{subfigure} 
    \caption{\justifying{\textbf{Spectral dispersion of the complex longitudinal Kerr angle} for s-polarized incident light ($\Phi_K^s$), illustrating Kerr rotation $\theta_K^s$ (dark green) and Kerr ellipticity $\eta_K^s$ (purple). Solid curves represent the total analytical response, while dash-dotted curves denote the MD reference response. \textbf{(a)} Terahertz domain. The Kaplan-Kittel exchange resonance at $\approx 0.0107$~eV ($2.6$~THz) emerges exclusively as a sharp peak in the Kerr rotation $\theta_K^s$, reflecting the phase-shifted nature of interfacial reflection. \textbf{(b)} Near-infrared domain. The additive coupling of the MD and ED mechanisms generates a distinct local extremum in the ellipticity $\eta_K^s$ and the zero value for $\theta_K^{s}$. The peak of this dome aligns with the vertical dashed line, denoting the crossover frequency where the magnitude of the decaying MD mechanism is perfectly compensated by the rising ED background ($|\mathrm{MD}| = |\mathrm{ED}|$).}}
    \label{fig:longitudinal_spectra}
\end{figure*}

\subsection{Longitudinal Kerr effect}\label{longitudinal_kerr}

The transverse Kerr geometry separates the electric- and magnetic-dipole contributions for the two orthogonal polarization channels. The longitudinal Kerr geometry provides the complementary case: the two contributions enter the same polarization-conversion amplitude and therefore interfere directly. This makes longitudinal MOKE particularly sensitive to the spectral competition between $\varepsilon_{xy}$ and $\mu_{xy}$ in the THz--NIR range.

In the longitudinal Kerr configuration (Fig.~\ref{fig:right}), the magnetization is oriented along the $z$-axis. To first order in the off-diagonal tensor elements, the two eigenmodes of the transmitted field are described by the refraction vector $\mathbf{n}_t^{\pm} = (n_{t,x}^{\pm}, 0, n_{t,z})$, where
\begin{equation}
	(n_{t,x}^{\pm})^2 = \varepsilon_{xx}\mu_{xx}-n_z^2\pm i \sqrt{\varepsilon_{xx}\mu_{xx}}n_z(\epsilon+\epsilon_\mu).
    \label{eq:n_long}
\end{equation}
Unlike the transverse configuration, the two eigenmodes are not purely $s$- or $p$-polarized. Evaluating $\hat{\mu} \mathbf{H}=\mathbf{n}_t^\pm\times \mathbf{E}$ and $\hat{\varepsilon}\mathbf{E} = -\mathbf{n}_t^\pm\times \mathbf{H}$ gives the corresponding mixed-polarization eigenvectors shown in Fig.~\ref{fig:right}.

The resulting polarization-conversion and diagonal reflection amplitudes are
\begin{align}
    r_{ps} = \frac{-\varepsilon_{xx}\mu_{xx}n_z\cos\varphi (\epsilon+\epsilon_\mu) }{\sqrt{\varepsilon_{xx}\mu_{xx}-n_z^2}(\mu_{xx}\cos\varphi+\sqrt{\varepsilon_{xx}\mu_{xx}-n_z^2})} \nonumber \\\cdot\frac{1}{(\varepsilon_{xx}\cos\varphi + \sqrt{\varepsilon_{xx}\mu_{xx}-n_z^2})} = -r_{sp},\label{eq:longitudinal_rps}
\end{align}
\begin{equation}
    r_{ss} = \frac{\mu_{xx}\cos\varphi-\sqrt{\varepsilon_{xx}\mu_{xx}-n_z^2}}{\mu_{xx}\cos\varphi+\sqrt{\varepsilon_{xx}\mu_{xx}-n_z^2}},
\end{equation}
\begin{equation}
    r_{pp} = \frac{\varepsilon_{xx}\cos\varphi-\sqrt{\varepsilon_{xx}\mu_{xx}-n_z^2}}{\varepsilon_{xx}\cos\varphi+\sqrt{\varepsilon_{xx}\mu_{xx}-n_z^2}}.
\end{equation}

For incidence from vacuum, the tangential wave-vector component is conserved, giving $n_z=\sin\varphi$. The longitudinal Kerr response for $s$-polarized incidence is then described by the complex Kerr angle
\begin{equation}
    \Phi_K^s = \theta_K^s + i\eta_K^s = \frac{r_{ps}}{r_{ss}}.
\end{equation}

The key result follows directly from Eq.~(\ref{eq:longitudinal_rps}): the polarization-conversion amplitude is proportional to
\begin{equation}
\epsilon+\epsilon_\mu
=
\frac{\varepsilon_{xy}}{\varepsilon_{xx}}
+
\frac{\mu_{xy}}{\mu_{xx}}.
\end{equation}
Thus, in contrast to the $s$-polarized transverse Kerr response, the longitudinal Kerr effect does not isolate the gyromagnetic contribution. The ED and MD mechanisms enter the same reflection channel and can therefore add or compensate spectrally. This provides a direct probe of their relative phase and magnitude.

Figure~\ref{fig:longitudinal_spectra} shows the resulting spectral dependence of $\theta_K^s$ and $\eta_K^s$. In the THz range [Fig.~\ref{fig:longitudinal_spectra}(a)], the ED contribution is weak and the response is dominated by $\mu_{xy}$. The Kaplan--Kittel resonance at $\hbar\omega\approx0.0107$~eV ($\nu\approx2.6$~THz) therefore produces a narrow feature in the Kerr rotation. The location of this resonance is identical to that obtained from the Faraday, polar-Kerr, and transverse-Kerr calculations because all four geometries use the same frequency-dependent permeability tensor. The different appearance of the resonance in the Kerr rotation, rather than in the ellipticity, results from the complex phase relation imposed by reflection and does not represent a change in the underlying magnetic resonance.

The NIR range provides the more significant result for the longitudinal geometry [Fig.~\ref{fig:longitudinal_spectra}(b)]. Here, the MD contribution decreases with increasing photon energy, whereas the dispersive ED contribution increases toward the high-energy electronic resonances. Because both terms enter $r_{ps}$ through the same combination $\epsilon+\epsilon_\mu$, their interference produces a pronounced extremum in the Kerr ellipticity. The position of this extremum coincides with the condition $
|\mathrm{MD}|=|\mathrm{ED}|$,
identifying the point at which the two mechanisms become comparable. Thus, unlike the Faraday and transverse-Kerr geometries, where the relative MD and ED contributions can be separated directly through the measured polarization channels, the longitudinal Kerr effect converts their spectral competition into a characteristic extremum of a single observable.

These findings make the longitudinal geometry particularly useful for locating the MD--ED crossover in the NIR.

\begin{figure*}[htb!]
    \centering
\begin{subfigure}[t]{0.47\textwidth}
				\centering
				\includegraphics[width=\textwidth]{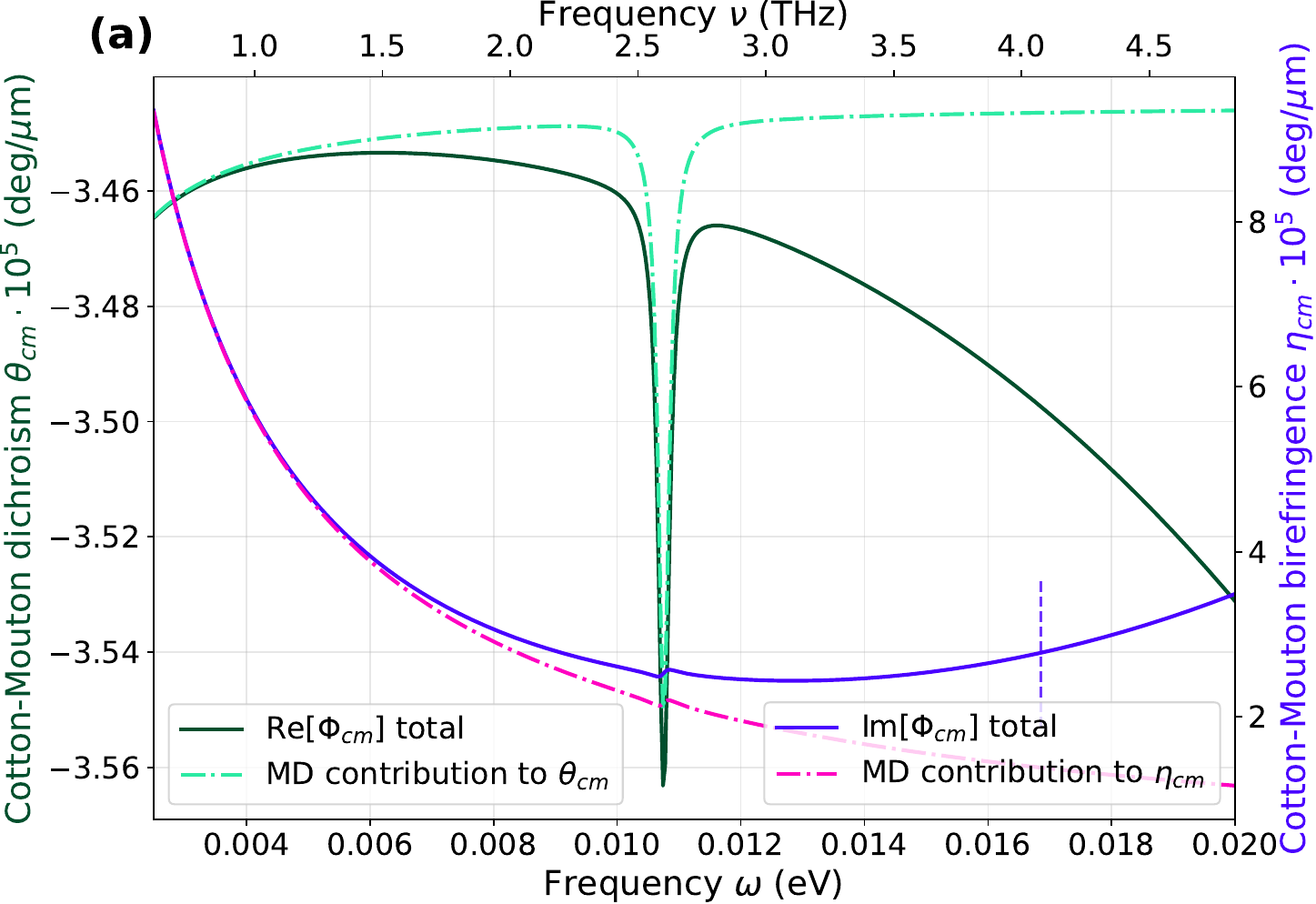}
				\label{fig:cm_thz}
			\end{subfigure}
			\begin{subfigure}[t]{0.47\textwidth}
				\centering
				\includegraphics[width=\textwidth]{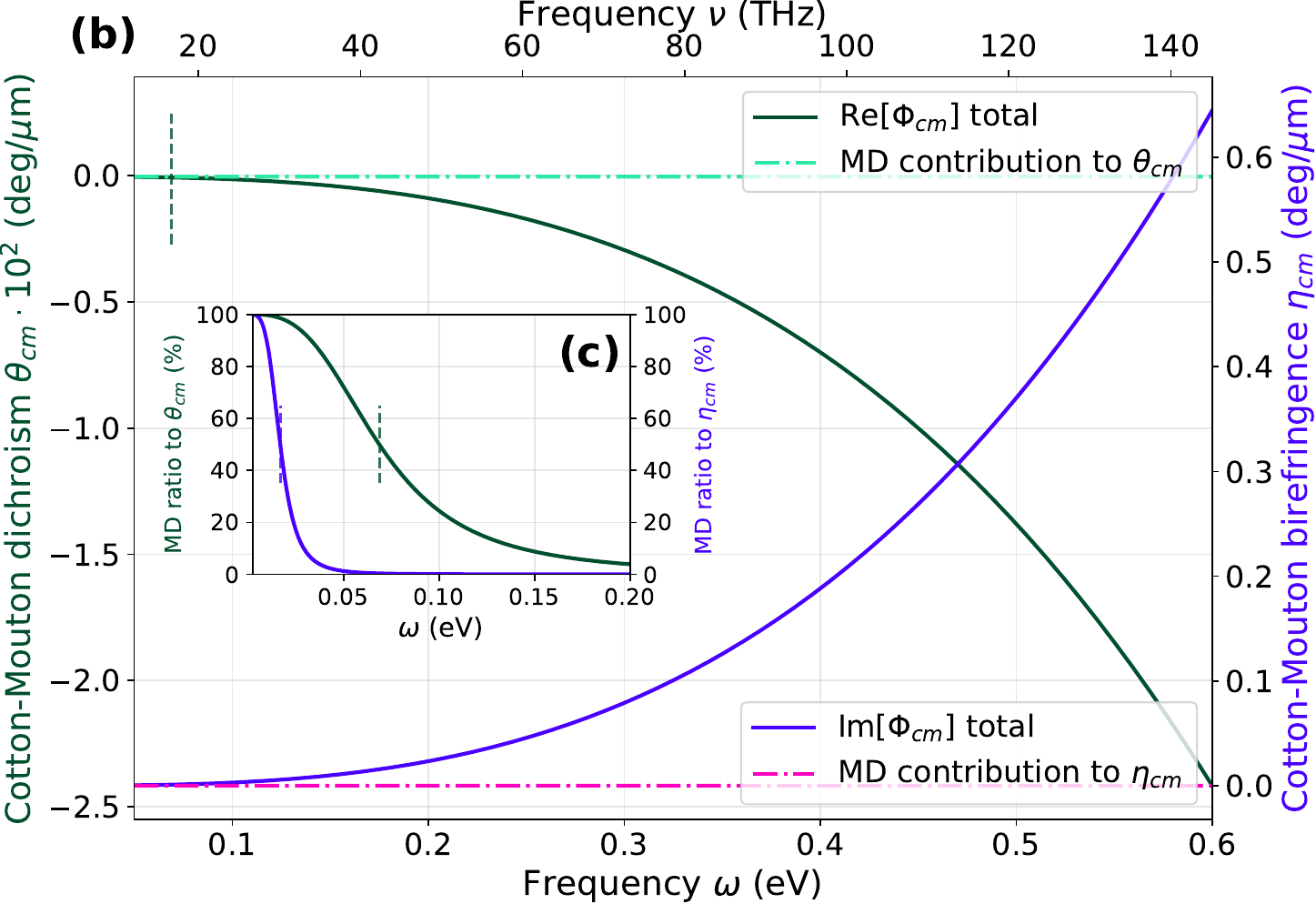}
				\label{fig:cm_nir}
			\end{subfigure} 
    \caption{\justifying{\textbf{Spectral dispersion of the Cotton-Mouton effect}: magnetic linear dichroism $\theta_{cm}$ (dark green curves) and birefringence $\eta_{cm}$ (purple curves). Solid lines represent the total analytical solutions, while dash-dotted lines isolate the MD contribution. \textbf{(a)} Terahertz domain. The MD mechanism fully dominates, enabling the predicted spectral signature of the exchange magnon resonance peak at $\nu \approx 2.6$~THz ($0.0107$~eV) in both optical parameters. The vertical dashed line marks the exact MD-ED crossover point for $\eta_{cm}$. \textbf{(b)} Near-infrared domain, demonstrating the spectral crossover between the decaying MD mechanism and the rising ED background. \textbf{(c)} Inset quantifying the fractional ratio of the MD reference response relative to the total effect. Vertical dashed lines denote the frequencies, where the MD-reference response reaches 50\% of the full response.}}
    \label{fig:cm}
\end{figure*}

\subsection{Magnetic linear birefringence and magnetic linear dichroism}\label{cotton_mouton}

The previous sections show that the contribution of the magnetic permeability becomes essential in the THz--NIR magneto-optical response and that its manifestation depends strongly on the experimental geometry. The Cotton--Mouton effect provides a complementary transmission geometry in which the two orthogonal linear polarizations acquire different complex refractive indices. The resulting magnetic linear birefringence and dichroism therefore offer a direct test of whether the frequency-dependent magnetic response associated with $\mu_{xy}$ survives outside the circular-polarization channels considered above.

The Cotton--Mouton effect is the magnetically induced difference between the optical responses for polarizations parallel and perpendicular to the magnetization direction \cite{zvezdin1997modern,sato2022fundamentals}. For linearly polarized light propagating perpendicular to $\mathbf M$, the two polarization components acquire different complex refractive indices, producing both magnetic linear birefringence and magnetic linear dichroism. In ferrimagnetic garnets, these effects have previously been studied experimentally at optical and NIR wavelengths and described using phenomenological and two-sublattice models \cite{smolensky1978birefringence,ando1987birefringence,sato1989cotton}. The present calculation extends this analysis to the frequency-dependent tensorial regime in which both $\hat{\varepsilon}(\omega)$ and $\hat{\mu}(\omega)$ are retained.

The magnetic linear dichroism $\theta_{cm}$ and magnetic linear birefringence $\eta_{cm}$ are combined into the complex specific Cotton--Mouton response
\begin{equation}
	\Phi_{cm} = \theta_{cm}+i\eta_{cm}=-\frac{i\pi }{\lambda}(n_t^s-n_t^p),\label{eq:Cotton-Mouton}
\end{equation}
where $n_t^s$ and $n_t^p$ are derived in Section~\ref{transverse_kerr}. The two quantities originate from the imaginary and real parts, respectively, of the difference between the complex refractive indices. In an absorbing medium, the differential attenuation of the two linear polarization components produces magnetic linear dichroism, whereas their differential phase accumulation produces magnetic linear birefringence.

For the present BiYIG parameters, the Cotton--Mouton response provides a direct measure of the competition between the electric and magnetic contributions already identified in the Faraday and Kerr geometries. The THz and NIR spectra are therefore evaluated using the same $\varepsilon_{ij}(\omega)$ and $\mu_{ij}(\omega)$ without introducing additional material parameters. This is important because previous studies of linear birefringence in garnets have predominantly interpreted the effect in terms of optical electronic transitions and phenomenological magneto-optical coefficients \cite{smolensky1978birefringence,ando1987birefringence,sato1989cotton}. In contrast, the present treatment explicitly retains the dynamic magnetic permeability, allowing the contribution associated with ferrimagnetic precession and the Kaplan--Kittel mode to be separated from the electronic response.

Figure~\ref{fig:cm}(a) shows the THz response. The calculated magnetic linear birefringence and dichroism are dominated by the MD contribution over the spectral range considered. A narrow feature occurs at $\hbar\omega\approx0.0107$~eV ($\nu\approx2.6$~THz), coinciding with the Kaplan--Kittel resonance obtained independently from the two-sublattice permeability tensor. Because the oscillator strength of this mode is strongly suppressed by the near equality of the sublattice $g$-factors, its dissipative contribution remains confined to a narrow resonance interval, while the surrounding response is determined by the dispersive magnetic permeability. Thus, the THz Cotton--Mouton response provides an additional optical signature of the same high-frequency ferrimagnetic dynamics identified in the Faraday and Kerr geometries. The result is consistent with the broader use of THz magnetic linear birefringence and dichroism as probes of collective spin dynamics in magnetic materials \cite{baierl2016thz}.

The significance of the THz result is not that the Cotton--Mouton effect itself is absent from the previous literature. THz magnetic linear birefringence and dichroism have already been used to detect spin dynamics in antiferromagnetic materials \cite{baierl2016thz}, while linear birefringence in ferrimagnetic garnets has long been interpreted within two-sublattice models \cite{smolensky1978birefringence}. Rather, the present calculation identifies the physical origin of the THz Cotton--Mouton response in BiYIG by explicitly tracing it to the frequency-dependent off-diagonal magnetic permeability and, in particular, to the Kaplan--Kittel branch of the two-sublattice response. This separates the magnetic contribution from the electronic contribution at the level of the constitutive tensors rather than through a phenomenological fit of the final magneto-optical signal.

The NIR response [Fig.~\ref{fig:cm}(b)] provides the complementary crossover regime. As the photon energy increases, the MD contribution decreases away from the low-frequency magnetic resonances, whereas the ED contribution increases owing to the dispersive tails of the electronic transitions contained in $\varepsilon_{xy}(\omega)$. The relative contributions therefore evolve continuously from an MD-dominated response toward an ED-dominated regime. The inset in Fig.~\ref{fig:cm}(c) shows this evolution directly through the fraction of the total Cotton--Mouton signal arising from the MD contribution. The MD-reference response reaches 50\% of the full response at different energies for the two observables: $\eta_{cm}$ crosses into the ED-dominated regime at $\hbar\omega\approx0.02$~eV, whereas the corresponding crossover for $\theta_{cm}$ occurs at $\hbar\omega\approx0.07$~eV. The different crossover energies reflect the different sensitivity of birefringence and dichroism to the dispersive and absorptive parts of the constitutive tensors.

The Cotton--Mouton results therefore provide an independent confirmation of the spectral hierarchy established in the preceding geometries. In the THz range, the response is controlled by the magnetic permeability and exhibits the Kaplan--Kittel resonance; in the NIR, the magnetic and electric contributions become comparable and eventually the electronic contribution dominates. This behavior is not a contradiction of earlier optical and infrared measurements of linear birefringence in garnets \cite{smolensky1978birefringence,ando1987birefringence,sato1989cotton}; rather, it extends their spectral analysis by resolving the magnetic-dipole contribution through the dynamic permeability tensor. The resulting separation of MD and ED mechanisms is the central advantage of the present tensorial treatment and establishes the Cotton--Mouton effect as a complementary probe of the high-frequency magnetic response of BiYIG.

\section*{Discussion}
The analytical results presented here correspond to the homogeneous-medium/single-interface limit. The same constitutive tensors can be incorporated into the finite-thickness 4×4 transfer-matrix formalism given in Appendix~\ref{app:tmm}, allowing direct extension to thin layers and multilayer structures.

The results establish that the broadband magneto-optical response of bismuth-substituted iron garnets cannot be characterized by the dielectric tensor alone once the photon energy is reduced from the electronic-resonance region toward the THz range. The central result is the separation of the gyroelectric and gyromagnetic responses at the level of the constitutive tensors, $\varepsilon_{xy}(\omega)$ and $\mu_{xy}(\omega)$, and their subsequent propagation through different experimental geometries. Rather than producing five independent magneto-optical calculations, the tensorial treatment provides a common material response from which the distinct polarization and phase sensitivities of Faraday, polar Kerr, transverse Kerr, longitudinal Kerr, and Cotton--Mouton effects follow.

The calculated response identifies two physically distinct magnetic contributions. The Kaplan--Kittel mode produces the resonant magnetic response in the THz range, whereas the much larger oscillator strength of the acoustic ferrimagnetic mode determines the high-frequency tail of the permeability. This distinction is important for interpreting the infrared response: the observation of a magnetic-dipole contribution at photon energies well above the Kaplan--Kittel resonance should not be attributed to the exchange-mode resonance itself, but to the broadband magnetic-permeability response of the ferrimagnet. The resulting MD--ED crossover is therefore a competition between the electronic gyrotropic response encoded in $\varepsilon_{xy}$ and the high-frequency tail of the ferrimagnetic permeability encoded in $\mu_{xy}$. The narrow Kaplan--Kittel resonance and the broadband MD background should consequently be regarded as separate spectral signatures of the same two-sublattice magnetic dynamics.

The different optical geometries provide complementary access to these contributions. In transmission, the Faraday effect combines the gyroelectric and gyromagnetic terms in the circular eigenmode splitting. In reflection, the phase relation imposed by the interface redistributes the dispersive and dissipative parts between Kerr rotation and ellipticity, producing different crossover frequencies for the two observables. The transverse Kerr geometry provides the strongest polarization selectivity: for $s$-polarized incidence, the magnetization-dependent term contains $\mu_{xy}$ but not $\varepsilon_{xy}$. The calculated signal at $\varphi=45^\circ$ reaches approximately $10^{-5}$ in the THz range, compared with approximately $10^{-7}$ for the corresponding $p$-polarized response. This two-order-of-magnitude difference identifies $s$-polarized TMOKE as a particularly sensitive geometry for detecting the intrinsic gyromagnetic response. The longitudinal geometry provides the complementary limit in which the two mechanisms enter the same polarization-conversion channel through $\varepsilon_{xy}/\varepsilon_{xx}+\mu_{xy}/\mu_{xx}$, converting their spectral competition into an extremum of the Kerr ellipticity. Finally, the Cotton--Mouton response supplies an independent transmission geometry in which the difference between the $s$- and $p$-polarized refractive indices probes the same material tensors.

This geometry dependence has direct experimental significance. The magnetic contribution does not have a unique spectral cutoff: its apparent importance depends on the observable and on whether the measurement probes propagation, circular reflection, polarization-selective reflection, or linear birefringence. Consequently, omission of $\mu_{xy}$ based only on the photon energy is not sufficient. Instead, the appropriate criterion is whether the specific optical observable is sensitive to the gyromagnetic tensor component. In particular, the $s$-polarized transverse Kerr channel provides a direct test of $\mu_{xy}$ without an off-diagonal dielectric contribution, while longitudinal Kerr and Faraday geometries reveal the interference between the two mechanisms. The resulting set of observables therefore provides multiple experimentally distinguishable signatures of the same underlying magnetic response.

The present results also place earlier magneto-optical studies of garnets in a common electrodynamic framework. Infrared Faraday measurements established the importance of magnetic-dipole and exchange-related contributions in rare-earth iron garnets decades ago \cite{krinchik1960faraday,krinchik1961infrared,krinchik1962exchange,crossley1969faraday}. Linear birefringence in YIG and Bi-substituted iron garnets was likewise established experimentally at optical and near-infrared wavelengths \cite{smolensky1978birefringence,ando1987birefringence}. More recently, gyromagnetic Kerr response has been demonstrated in an engineered magneto-plasmonic metamaterial \cite{yang2022gyromagnetic}. The present results extend these observations by treating the intrinsic magnetic permeability of a ferrimagnetic garnet on the same footing as its dielectric response and by following the resulting MD contribution continuously from the THz magnetic resonance into the infrared regime.

Several limitations define the quantitative scope of the predictions. The dielectric tensor below the experimentally characterized optical range is obtained from a resonant extrapolation, so the calculated crossover frequencies depend on the accuracy of this low-energy continuation. Independent infrared optical or magneto-optical data would therefore provide an important validation of the extrapolated $\varepsilon_{ij}(\omega)$. The magnetic response likewise depends on the adopted damping parameters and on the exchange-field parameter determining the Kaplan--Kittel frequency. These quantities should be specified and, where possible, constrained by independent measurements. Finally, the present Fresnel treatment describes a homogeneous magnetic medium; quantitative comparison with a finite Bi:YIG film on a substrate requires the corresponding multilayer optical boundary problem. Within these assumptions, however, the calculated polarization-selective signatures provide experimentally testable predictions for separating magnetic-permeability and dielectric-permittivity contributions in the THz--infrared magneto-optics of ferrimagnetic garnets.

        \section*{Conclusion}

A frequency-dependent tensorial description of the broadband magneto-optical response of bismuth-substituted iron garnets has been developed by treating the dielectric permittivity and magnetic permeability on equal footing. The electric-dipole contribution is determined by the off-diagonal dielectric response $\varepsilon_{xy}(\omega)$, whereas the magnetic-dipole contribution is determined by the dynamic permeability $\mu_{xy}(\omega)$ obtained from the two-sublattice ferrimagnetic response. This common material description is applied without changing the material parameters to Faraday, polar Kerr, transverse Kerr, longitudinal Kerr, and Cotton--Mouton geometries.

The calculations predict a magnetic-dipole-dominated magneto-optical response in the THz range, with the Kaplan--Kittel exchange mode producing a narrow resonant signature at the calculated Bi:YIG frequency. At higher infrared frequencies, the magnetic contribution evolves into a broadband permeability tail and competes with the electronic gyrotropic response, resulting in geometry- and observable-dependent MD--ED crossover frequencies. The $s$-polarized transverse Kerr effect is particularly selective because its magnetization-dependent contribution is controlled by $\mu_{xy}$ without an off-diagonal dielectric term; at $\varphi=45^\circ$, the calculated THz signal is approximately two orders of magnitude larger than the corresponding $p$-polarized response. In the longitudinal Kerr geometry, the two mechanisms instead enter the same polarization-conversion channel, producing a characteristic extremum at their spectral compensation. The Cotton--Mouton effect provides an independent transmission signature of the same magnetic response.

These results show that the magnetic contribution to long-wavelength magneto-optics cannot be represented by a single high-frequency cutoff. Its observability depends on both the spectral dependence of the material tensors and the symmetry of the optical geometry. The resulting tensorial framework provides experimentally testable routes for isolating the intrinsic gyromagnetic response of ferrimagnetic garnets and for probing high-frequency exchange dynamics through polarization-selective THz and infrared magneto-optics.

\section*{Acknowledgments} 
The author is grateful to A.N. Kalish for stimulating discussions and valuable insights.

	\begin{widetext}

		\appendix
		
\section{Dielectric permittivity tensor}\label{tensor_eps}

\begin{figure*}[htb!]
    \centering
\includegraphics[width=0.7\textwidth]{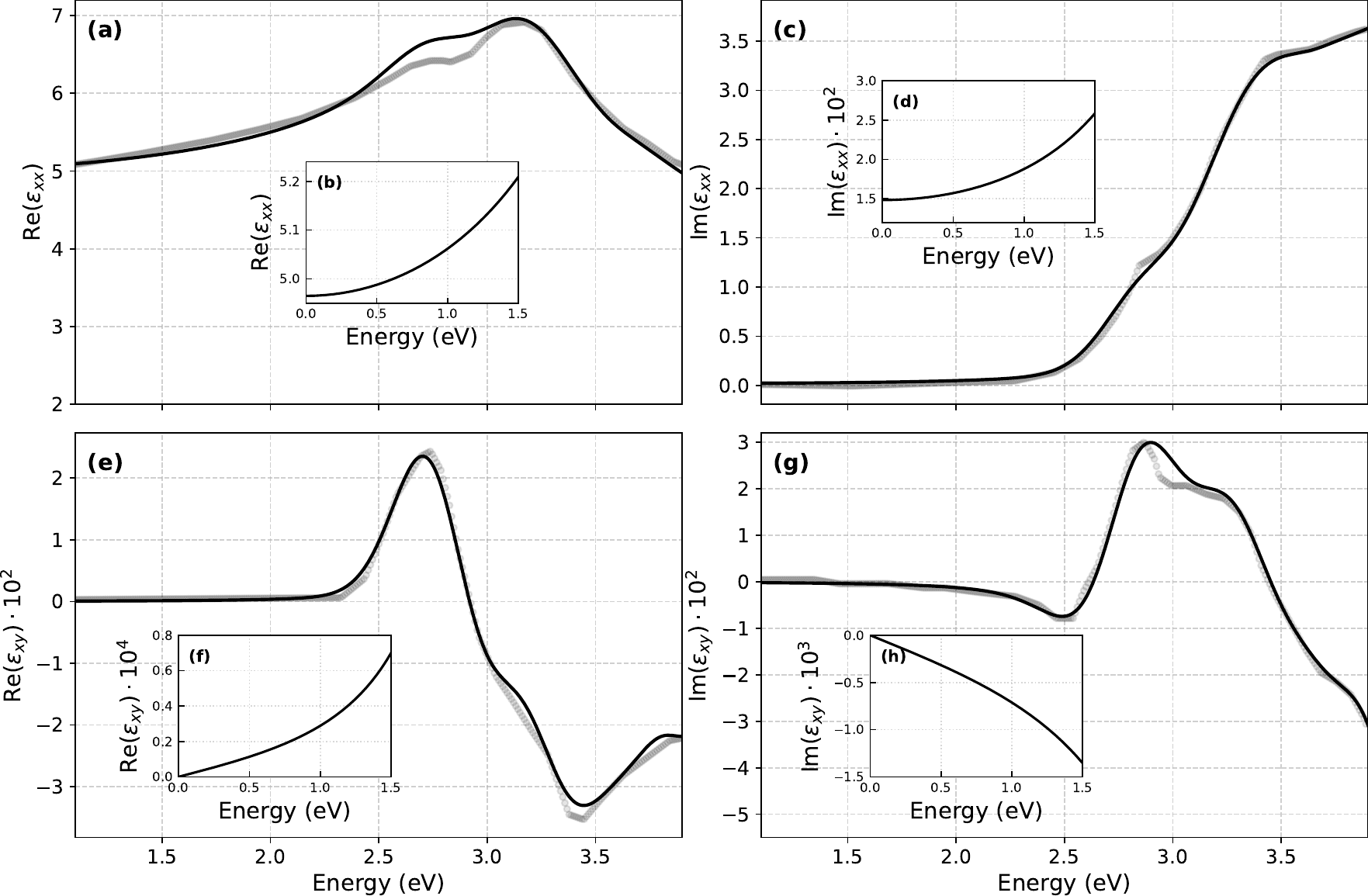}
				\label{fig:epsilon}
    \caption{\justifying{Real and imaginary parts of the diagonal (a, c) and off-diagonal (e, g) components of the dielectric tensor at concentration $x=0.2$. Semi-transparent black dots denote the experimental data from Ref.~\cite{krumme1989optical}. The solid black curves in both the main panels and the insets represent the total fit obtained by summing all resonances over the visible range. The insets (b, d, f, h) display the low-energy extrapolation of the total fit. Accurately fitting the entire visible spectrum is essential for properly evaluating the low-frequency behavior of the dielectric tensor.}}
    \label{fig:epsilon}
\end{figure*}
Magneto-optical responses in magnetic materials are fundamentally governed by two distinct types of optical transitions: diamagnetic and paramagnetic. Diamagnetic transitions arise from the Zeeman splitting of the excited or ground states, maintaining equal oscillator strengths for left- and right-circularly polarized light. This fundamental symmetry manifests as a characteristic derivative-like lineshape in the off-diagonal components of the dielectric tensor. Conversely, paramagnetic transitions originate either from a difference in the thermal populations of the Zeeman-split ground state sublevels or from intrinsically unequal transition probabilities for orthogonal circular polarizations. Consequently, paramagnetic contributions produce distinct absorptive lineshapes, dominating the magneto-optical spectra at low temperatures or high magnetic fields where polarization of the electronic spins is maximized.
		
		To formally describe these excitations, we introduce the diagonal, $\varepsilon_{xx}$, and off-diagonal, $\varepsilon_{xy}$, components of the dielectric tensor. For a purely homogeneous system, the diamagnetic contribution is expressed as
		\begin{align}
			\varepsilon_{xx}^{(d)} &= \frac{K_0}{2} \sum_{j=1,2} \sum_{s=\pm} t_{js}(\omega), \\ \varepsilon_{xy}^{(d)} &= i\frac{K_0}{2} \sum_{j=1,2} (-1)^{j+1} \sum_{s=\pm} s \, t_{js}(\omega),
		\end{align}
		where the bare Lorentzian poles are defined as $t_{1\pm}(\omega) = (\omega_0 + \Delta \pm \omega - i\Gamma)^{-1}$ and $t_{2\pm}(\omega) = (\omega_0 - \Delta \pm \omega - i\Gamma)^{-1}$. Here, $K_0$ denotes the effective amplitude proportional to the transition oscillator strength, $\omega_0$ is the resonance frequency, $\Gamma$ is the homogeneous damping rate, and $\Delta$ represents the energy shift induced by the effective exchange or Zeeman splitting. Depending on the relative signs of the band-specific Land\'{e} factors and exchange integrals, $\Delta$ can inherently take negative values. Analogously, the paramagnetic tensor components are given by
		\begin{align}
			\varepsilon_{xx}^{(p)} &= \sum_{j=1,2} \frac{K_j}{2} \sum_{s=\pm} t_{js}(\omega), \\ \varepsilon_{xy}^{(p)} &= i\sum_{j=1,2} (-1)^{j} \frac{K_j}{2} \sum_{s=\pm} s \, t_{js}(\omega),
		\end{align}
		where $K_1$ and $K_2$ account for the unequal oscillator strengths of the respective split transitions. The microscopic derivation of these tensor forms from the fundamental Kubo formula is provided in the Appendix.
		
		While the purely Lorentzian description captures the core resonance physics, it inherently overestimates the spectral weight at the extended band tails and frequently fails to resolve the low-energy discrepancies observed in experimental spectra. Real materials exhibit local structural disorder and compositional fluctuations that require the incorporation of inhomogeneous broadening. This is rigorously achieved by convolving the Lorentzian poles with a Gaussian distribution, naturally yielding the Voigt profile. Mathematically, the bare poles $1/D_{js}$ are substituted using the complex Faddeeva function $w(z) = e^{-z^2}\mathrm{erfc}(-iz)$:
		\begin{equation}
			V(D_{js}, \sigma) = i\frac{\sqrt{\pi/2}}{\sigma} w\left( \frac{-D_{js}}{\sigma \sqrt{2}} \right),
		\end{equation}
		where $\sigma$ parameterizes the Gaussian standard deviation of the inhomogeneous broadening, and $D_{js}$ represents the complex denominators of $t_{js}(\omega)$. The dielectric tensor components are exactly reformulated by replacing every instance of $t_{js}(\omega)$ with $V(D_{js}, \sigma)$, providing a robust phenomenological model that suppresses unphysical low-energy tails and accurately captures the lineshape asymmetry.
		
		The macroscopic optical response of the system is thus formulated as the linear superposition of a high-frequency background permittivity and all isolated resonances:
		\begin{equation}
			\varepsilon_{ij}(\omega) = \varepsilon_{ij}^{(\infty)} \delta_{ij} + \sum_{n} \varepsilon_{ij}^{(d, n)}(\omega) + \sum_{m} \varepsilon_{ij}^{(p, m)}(\omega),
		\end{equation}
		where $\varepsilon^{(\infty)}$ is the purely diagonal non-resonant background, and the sums iterate over all active diamagnetic and paramagnetic transitions, respectively.
		
		To extract the physical transition parameters from the experimental datasets, a hybrid global-local nonlinear optimization was employed simultaneously across multiple dopant concentrations ($x=0.2, 0.4, 0.6, 0.8, 1.0, 1.2, 1.4$). But for this work the bismuth consentration $x=0.2$ was selected, the dielectric permittivity tensor is presented and used in the Figure~\ref{fig:epsilon}. The procedure minimizes the soft-$\ell_1$ loss function of the residuals for both the real and imaginary parts of $\varepsilon_{xx}$ and $\varepsilon_{xy}$, ensuring resilience against experimental noise and localized spectral outliers. A critical constraint in this model is the enforcement of a globally shared, concentration-independent inhomogeneous broadening parameter, $\sigma$. Fixing $\sigma$ as a global variable prevents non-physical over-parameterization and maintains mathematical consistency, under the assumption that the fundamental defect-induced disorder remains comparable across the studied concentration range. The resonance frequencies $\omega_0$ were constrained within narrow spectral windows ($\pm 0.15$ eV from initial structural estimates) to rigidly preserve their physical assignment to specific interband transitions. Furthermore, to reflect the underlying mean-field physics of dilute magnetic semiconductors, the optimization boundary conditions explicitly permit the splitting parameter $\Delta$ in diamagnetic transitions to grow systematically with increasing concentration. It should be explicitly noted that the exact atomic concentration $x$ cannot be independently determined solely from these magneto-optical fits due to the highly correlated scaling between the effective transition amplitudes and local structural perturbations; therefore, $x$ acts as an externally assigned parameter linked to independent macroscopic compositional analysis.
		
		The resonant representation used here is causal: each damped Lorentz pole is analytic in the upper half of the complex-frequency plane, and the Voigt profile is obtained by averaging these causal responses over a real Gaussian distribution of resonance frequencies. The resulting dielectric functions therefore satisfy the corresponding Kramers--Kronig constraints. The low-energy continuation is nevertheless model dependent because the experimental data constrain the fit primarily in the visible spectral range. The purpose of the present parametrization is not to establish a unique reconstruction of $\varepsilon_{ij}$ at low photon energies, but to provide a causal continuation of the measured optical response for evaluating the magneto-optical effects considered below.

\section{Magnetic Permeability Tensor}\label{app:mu_tensor}

The dynamic magnetic permeability tensor $\hat{\mu}$ for bismuth-substituted yttrium iron garnet (BiYIG) is modeled using a rigorous two-sublattice ferrimagnetic framework. The crystal contains $\text{Fe}^{3+}$ magnetic ions distributed across tetrahedral ($d$) and octahedral ($a$) sites, characterized by the macroscopic sublattice magnetizations $M_1$ and $M_2$, and distinct experimental Landé factors $g_1$ and $g_2$. The permeability tensor components $\mu_{xx}$ and $\mu_{xy}$ are derived from the coupled Landau-Lifshitz-Gilbert equations for the two sublattices and are expressed as a sum over two normal precessional modes ($k=1,2$):
\begin{equation}
	\mu_{xx} = 1 + \sum_{k=1,2} \omega_{Mk} \dfrac{ \omega_k \left[ \omega_k^2-\left(1-\alpha_k^2\right) \omega^2 \right]-i\alpha_k \omega\left[\omega_k^2+\left(1+\alpha_k^2\right) \omega^2\right]}{\left[\omega_k^2-\left(1+\alpha_k^2\right) \omega^2\right]^2+4 \alpha_k^2 \omega^2 \omega_k^2},
	\label{eq:muxx_ferri}
\end{equation}
\begin{equation}
	\mu_{xy} = \sum_{k=1,2} \omega_{Mk} \dfrac{i \omega\left[\omega_k^2-\left(1+\alpha_k^2\right) \omega^2\right] + 2 \alpha_k \omega^2\omega_k}{\left[\omega_k^2-\left(1+\alpha_k^2\right) \omega^2\right]^2+4 \alpha_k^2 \omega^2 \omega_k^2}.
	\label{eq:muxy_ferri}
\end{equation}

The fundamental parameters governing these modes are based on the saturation magnetization $4\pi M_{net} = 1830$~G and an applied magnetic field $H_0 = 4000$~Oe. The sublattice magnetizations follow the established 3:2 site distribution ratio for BiYIG, yielding $M_1 = 3M_{net}$ and $M_2 = 2M_{net}$. The gyromagnetic ratios for the sublattices are defined as $\gamma_j = g_j \gamma_0$, where $\gamma_0 \approx 8.8 \times 10^{-6}$~$\text{ps}^{-1}\text{Oe}^{-1}$ is the base gyromagnetic ratio, and the experimentally determined $g$-factors are $g_1 = 2.005$ and $g_2 = 2.003$.

The $k=1$ term corresponds to the acoustic ferromagnetic resonance (FMR) mode, representing the collective, in-phase precession of the coupled sublattices. Its resonance frequency and effective oscillator strength are given by:
\begin{align}
    \omega_1 &= \gamma_{eff} H_0, \\
    \omega_{M1} &= 4\pi \gamma_{eff} (M_1 - M_2),
\end{align}
where $\gamma_{eff} = (M_1 - M_2)/(M_1/\gamma_1 - M_2/\gamma_2)$ is the effective gyromagnetic ratio. Numerical evaluation yields a resonance frequency of $\hbar\omega_1 \approx 4.63 \times 10^{-5}$~eV (equivalent to $11.2$~GHz) and an oscillator strength of $\hbar\omega_{M1} \approx 2.12 \times 10^{-5}$~eV.

The $k=2$ term represents the Kaplan-Kittel Kaplan-Kittel mode, an optical magnon branch characterized by the out-of-phase precession of the sublattice magnetizations, which is strongly driven by the intersublattice exchange interaction \cite{kaplan1953exchange}.In the two-sublattice ferrimagnetic model, the Kaplan-Kittel exchange mode frequency $\omega_2$ is proportional to the exchange field between the sublattices. In the limit where the exchange field vastly exceeds the applied field ($H_{ex} \gg H_0$), the analytical expression for the Kaplan-Kittel resonance frequency is:
\begin{equation}
\omega_2 \approx \lambda (\gamma_2 M_1 - \gamma_1 M_2)
\end{equation}

For BiYIG, characterized by a dimensionless exchange constant $\lambda \approx 6.38 \times 10^3$, this exchange-dominated mode lies deep in the terahertz domain at $\nu_{ex} = \omega_2/2\pi = 2.6$~THz, establishing the resonance frequency $\hbar\omega_2 \approx 0.0107$~eV. Its oscillator strength is strictly defined as:
\begin{equation}
    \omega_{M2} = 4\pi \frac{M_1 M_2 (\gamma_1 - \gamma_2)^2}{\gamma_2 M_1 - \gamma_1 M_2}.
\end{equation}
Calculating this expression yields an exceptionally small value of $\hbar\omega_{M2} \approx 1.27 \times 10^{-10}$~eV. 

A comparative analysis of the oscillator strengths reveals that the low-frequency FMR mode is more than five orders of magnitude stronger than the high-frequency Kaplan-Kittel mode ($\omega_{M1} \gg \omega_{M2}$). This vast disparity originates from the physical symmetry of the precessional dynamics. The FMR oscillator strength ($\omega_{M1}$) is proportional to the net macroscopic magnetization $(M_1 - M_2)$. Conversely, the Kaplan-Kittel mode represents an antiferromagnetic-like excitation; its effective macroscopic magnetization ($\omega_{M2}$) scales with the squared difference of the respective gyromagnetic ratios $(\gamma_1 - \gamma_2)^2$. Because the local crystalline environments of the $\text{Fe}^{3+}$ ions in the $d$ and $a$ sites generate nearly identical Landé factors ($g_1 \approx g_2$), this difference is vanishingly small, severely suppressing the coupling of the optical mode to the external electromagnetic field. 

The Kaplan--Kittel mode governs the resonant magnetic response in the THz range, whereas the much larger oscillator strength of the acoustic mode determines the broadband high-frequency tail of the permeability. Thus, the THz resonance and the infrared MD background originate from distinct parts of the same two-sublattice magnetic response.
A uniform phenomenological Gilbert damping parameter of $\alpha_1 = \alpha_2 = 0.01$ is adopted across both precessional regimes.

\begin{figure*}[htb!]
			\centering
			\begin{subfigure}[t]{0.29\textwidth}
				\centering
				\includegraphics[width=\textwidth]{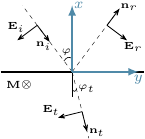}
				\caption{Transverse Kerr configuration}
				\label{fig:middle}
			\end{subfigure}
			\hspace{0.05\textwidth}
			\begin{subfigure}[t]{0.29\textwidth}
				\centering
				\includegraphics[width=\textwidth]{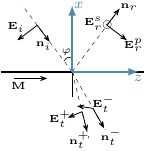}
				\caption{Longitudinal Kerr configuration}
				\label{fig:right}
			\end{subfigure}
			\caption{Kerr magneto-optical configurations for p-polarized incident light: (a) transverse and (b) longitudinal setups.}
			\label{fig:three-images}
		\end{figure*}

		\section{Electromagnetic field modes in the transverse Kerr geometry}

An incident electromagnetic $p$-wave (TM mode) is determined by $\mathbf{E}_i = -E_i (\sin\varphi, \cos\varphi, 0) e^{-i\omega t + i\frac{\omega}{c} \mathbf{n}_i \cdot \mathbf{r}}$ and $\mathbf{n}_i = n_1 (-\cos\varphi, \sin\varphi, 0)$,
		where $n_1=1$ in the case of free space. 
		The reflected electromagnetic field is also $p$-polarized and takes the form $\mathbf{E}_r = E_r (-\sin\varphi, \cos\varphi, 0) e^{-i\omega t + i\frac{\omega}{c} \mathbf{n}_r \cdot \mathbf{r}}$ and $\mathbf{n}_r = n_1 (\cos\varphi, \sin\varphi, 0)$.

		In the transverse Kerr effect configuration (Fig.~\ref{fig:middle}), the transmitted light preserves polarization and satisfies Maxwell's equation $\hat{\varepsilon} \mathbf{E}_t = -\mathbf{n}_t \times \mathbf{H}_t$ with $\mathbf{n}_t = n_t (-\cos\varphi_t, \sin\varphi_t, 0)$, therefore, reads
		\begin{equation}
			\mathbf{E}_t = E_t \left(-\sin\varphi_t + \frac{\varepsilon_{xy}}{\varepsilon_{xx} \cos\varphi_t}, -\cos\varphi_t, 0\right) e^{-i\omega t + i\frac{\omega}{c} \mathbf{n}_t^p \cdot \mathbf{r}}, 
		\end{equation}
		where $n_t^p\approx\sqrt{\varepsilon_{xx}\mu_{zz}}$ is the refraction vector module for p-wave in the anisotropic medium, considering contributions up to the first order of smallness, and is given in Eq.~\ref{eq:nt_p}.
		
		Using interface conditions for electromagnetic fields, one can derive the reflection (Eq~\ref{eq:transverse_rpp}) and transmission Fresnel coefficients as
		\begin{align}
			t_{pp}&= \frac{2\sqrt{\varepsilon_{xx}\mu_{zz}}\cos\varphi\cos\varphi_t}{\cos \varphi (\varepsilon_{xx}\cos\varphi_t-\varepsilon_{xy}\sin\varphi_t)+\sqrt{\varepsilon_{xx}\mu_{zz}}\cos^2\varphi_t}
		\end{align}
		where contributions up to the first order of smallness ($|\mu_{xy}/\mu_{xx}|\ll1, |\varepsilon_{xy}/\varepsilon_{xx}|\ll1$) are considered.

		In the case of $s$-wave (TE mode), the incident light is determined by $\mathbf{H}_i = -H_i (\sin\varphi, \cos\varphi, 0) e^{-i\omega t + i\frac{\omega}{c} \mathbf{n}_i \cdot \mathbf{r}}$ and $\mathbf{n}_i = n_1 (-\cos\varphi, \sin\varphi, 0)$. 
		The reflected electromagnetic field is also $s$-polarized and takes the form $\mathbf{H}_r = H_r (-\sin\varphi, \cos\varphi, 0) e^{-i\omega t + i\frac{\omega}{c} \mathbf{n}_r \cdot \mathbf{r}}$ and $\mathbf{n}_r = n_1 (\cos\varphi, \sin\varphi, 0)$.
		
		Using Maxwell's equation $\hat{\mu} \mathbf{H}_t = \mathbf{n}_t^s \times \mathbf{E}_t$ with $\mathbf{n}_t^s = n_t^s (-\cos\varphi_t, \sin\varphi_t, 0)$, transmitted light takes the form
		\begin{equation}
			\mathbf{H}_t = H_t \left(-\sin\varphi_t + \frac{\mu_{xy}}{\mu_{xx} \cos\varphi_t}, -\cos\varphi_t, 0\right) e^{-i\omega t + i\frac{\omega}{c} \mathbf{n}_t^s \cdot \mathbf{r}},
		\end{equation}
		where $n_t^s\approx\sqrt{\varepsilon_{zz}\mu_{xx}}$ (Eq.~\ref{eq:nt_s}). Considering contributions up to the first order of smallness $|\mu_{xy}/\mu_{xx}|\ll1, |\varepsilon_{xy}/\varepsilon_{xx}|\ll1$ and using interface conditions for electromagnetic fields, one can derive the reflection (Eq.~\ref{eq:transverse_rss}) and transmission Fresnel coefficients as
		\begin{align}
			t_{ss}= \frac{2\cos\varphi(\mu_{xx}\cos\varphi_t-\mu_{xy}\sin\varphi_t)}{\cos \varphi (\mu_{xx}\cos\varphi_t-\mu_{xy}\sin\varphi_t)+\sqrt{\varepsilon_{xx}\mu_{xx}}\cos^2\varphi_t}.
		\end{align}

These expressions explicitly separate the two gyrotropic channels of the transverse geometry. The $p$-polarized mode contains the off-diagonal dielectric response through $\varepsilon_{xy}$, whereas the $s$-polarized mode contains the off-diagonal magnetic response through $\mu_{xy}$. In the conventional approximation $\hat{\mu}\rightarrow\mathbf{1}$, the $\mu_{xy}$ term vanishes and the magnetization-dependent $s$-polarized contribution disappears. Conversely, retaining the full permeability tensor introduces an intrinsically magnetic contribution to the $s$-polarized reflection and transmission amplitudes. This separation is the basis for the polarization-selective MD response discussed in Sec.~\ref{transverse_kerr}.

\section{Electromagnetic eigenmodes in the longitudinal Kerr geometry}

In the longitudinal Kerr geometry (Fig.~\ref{fig:right}), the magnetization lies in the plane of incidence. Unlike the transverse configuration, the eigenmodes are not separated into independent $s$- and $p$-polarized waves. The electric- and magnetic-dipole gyrotropies therefore enter the same electromagnetic eigenmodes and contribute simultaneously to polarization conversion.

To first order in the off-diagonal tensor elements, the refraction vector is given in Section~\ref{longitudinal_kerr} in Eq.\ref{eq:n_long}.
The two solutions correspond to the two magnetically nondegenerate eigenmodes of the anisotropic medium. Because the eigenvalues depend on the sum $\epsilon+\epsilon_\mu$, both gyroelectric and gyromagnetic responses contribute already at the level of the propagation modes. The general solution for the system is then given by the linear combination of magnetic-field eigenvectors        
\begin{equation}
	\mathbf{H}_{t}^{\pm}=H_z \frac{1}{\sqrt{\varepsilon_{xx}\mu_{xx}-n_z^2}}
	\begin{pmatrix}
		-n_z\pm\frac{i\sqrt{\varepsilon_{xx}\mu_{xx}}}{ (\varepsilon_{xx}\mu_{xx}-n_z^2)}\bigl(\varepsilon_{xx}\mu_{xx}\epsilon_\mu-\frac{n_z^2}{2}(\epsilon+\epsilon_\mu)\big)\\
        \mp i \sqrt{\varepsilon_{xx}\mu_{xx}}\bigl(1\pm\frac{i n_z\sqrt{\varepsilon_{xx}\mu_{xx}}}{2(\varepsilon_{xx}\mu_{xx}-n_z^2)}(\epsilon+\epsilon_\mu)\bigr)
		\\
		\sqrt{\varepsilon_{xx}\mu_{xx}-n_z^2}
	\end{pmatrix},
\end{equation}
and the corresponding electric-field eigenvectors are
\begin{equation}
	\mathbf{E}_{t}^{\pm}=H_z \frac{\sqrt{\varepsilon_{xx}\mu_{xx}}}{\varepsilon_{xx}\sqrt{\varepsilon_{xx}\mu_{xx}-n_z^2}}
	\begin{pmatrix*}
		\mp i n_z -\epsilon\sqrt{\varepsilon_{xx}\mu_{xx}}+\frac{n_z^2\sqrt{\varepsilon_{xx}\mu_{xx}}}{2(\varepsilon_{xx}\mu_{xx}-n_z^2)}(\epsilon+\epsilon_\mu)\\
		\sqrt{\varepsilon_{xx}\mu_{xx}}\mp \frac{i n_z \varepsilon_{xx}\mu_{xx}}{2(\varepsilon_{xx}\mu_{xx}-n_z^2)}(\epsilon+\epsilon_\mu)\\
		\pm i \sqrt{\varepsilon_{xx}\mu_{xx}-n_z^2}\bigl(1\pm \frac{i\sqrt{\varepsilon_{xx}\mu_{xx}} n_z}{\varepsilon_{xx}\mu_{xx}-n_z^2}(\epsilon+\epsilon_\mu)\bigr)
	\end{pmatrix*}.
\end{equation}

        This result provides the microscopic electromagnetic basis for the longitudinal Kerr response discussed in Sec.~\ref{longitudinal_kerr}. In particular, the magnetic-dipole contribution enters through $\epsilon_\mu=\mu_{xy}/\mu_{xx}$ and therefore carries the frequency dependence of the two-sublattice permeability, including the acoustic and Kaplan--Kittel modes. The longitudinal geometry consequently does not suppress the magnetic response; instead, it combines the magnetic-dipole and electric-dipole channels within the same eigenmodes and makes their spectral interference directly observable.
\section{Finite-thickness transfer-matrix formulation}\label{app:tmm}

The analytical expressions used in the main text describe either bulk propagation through a homogeneous medium or reflection from a single vacuum--medium interface. The same frequency-dependent dielectric and magnetic-permeability tensors can be used without modification for a finite-thickness BiYIG layer and for arbitrary multilayer structures by retaining the backward-propagating modes and all internal reflections. This is conveniently formulated within the $4\times4$ Berreman transfer-matrix formalism \cite{berreman1972,zak1990universal,zak1991multilayer,atkinson1993firstorder,rogers2011mueller,zamani2016bigyrotropic}.

For a monochromatic field with time dependence $\exp(-i\omega t)$ and conserved in-plane wave vector $k_x$, define $k_0=\frac{\omega}{c},q=\frac{k_x}{k_0}$
and introduce the field vector
$\boldsymbol{\Psi}^T(z)=
(
E_x, E_y,H_x,H_y).$

After eliminating the longitudinal components $E_z$ and $H_z$ from Maxwell's equations, the field evolution in a homogeneous layer can be written as
\begin{equation}
\frac{d\boldsymbol{\Psi}}{dz}
=
i k_0\,\mathbf{B}\boldsymbol{\Psi}, \qquad \mathbf{B}=
\begin{pmatrix}
0 & 0 & -\mu_{xy} &
\mu_{xx}-q^2/\varepsilon_{zz}\\
0 & 0 & -\mu_{xx} & -\mu_{xy}\\
\varepsilon_{xy} &
-\varepsilon_{xx}+q^2/\mu_{zz} & 0 & 0\\
\varepsilon_{xx} & \varepsilon_{xy} & 0 & 0
\end{pmatrix}.
\label{eq:berreman}
\end{equation}
Equation~\ref{eq:berreman} intrinsically contains both gyroelectric ($\varepsilon_{xy}$) and gyromagnetic ($\mu_{xy}$) responses. Therefore, no additional magneto-optical parameters are required to model finite thicknesses.

The matrix $\mathbf{B}$ assumes the magnetization is oriented along a specific axis. In a general experiment, the sample or its magnetization might be rotated relative to the optical table (e.g., switching between polar, longitudinal, and transverse configurations). Instead of recalculating Maxwell's equations for every possible rotation, the light's plane of incidence is fixed in the laboratory frame. If a layer's principal axes are rotated, its constitutive tensors ($\hat{\varepsilon}_{\rm mat}$, $\hat{\mu}_{\rm mat}$) are transformed into the laboratory frame via a rotation matrix $\mathbf{R}$, such as $\hat{\varepsilon}_{\rm lab} = \mathbf{R}\hat{\varepsilon}_{\rm mat}\mathbf{R}^{-1}$ and $\hat{\mu}_{\rm lab} = \mathbf{R}\hat{\mu}_{\rm mat}\mathbf{R}^{-1}.$
This single transformation allows the exact same formalism to universally solve any Kerr or Faraday geometry.

For a homogeneous layer $j$ of thickness $d_j$, the propagation matrix $\mathbf{P}_j(d_j)$ is constructed from the eigenvectors $\mathbf{V}_j$ and eigenvalues $\mathbf{Q}_j$ of its Berreman matrix $\mathbf{B}_j$:
\begin{equation}
    \mathbf{P}_j(d_j) = \mathbf{V}_j \exp\!\left(i k_0\mathbf{Q}_j d_j\right) \mathbf{V}_j^{-1}.
\end{equation}
For an $N$-layer structure, the total propagation matrix is the ordered product $\mathbf{P} = \mathbf{P}_N \cdots \mathbf{P}_1$.

The analytical formulas in the main text assume a single, semi-infinite vacuum/BiYIG interface, meaning light enters the material and propagates infinitely without returning. Real experimental setups, however, involve finite structures (e.g., air/BiYIG/substrate). When transmitted light hits the BiYIG/substrate interface, it reflects backward, returning to the front surface to coherently interfere with the initial reflection. This Fabry--Pérot interference fundamentally alters both the amplitude and phase of the measured signals.

The TMM strictly captures this by matching continuous tangential fields across all interfaces simultaneously. Decomposing the incident medium into forward ($\mathbf{a}_{\rm in}$) and backward ($\mathbf{a}_{\rm r}$) modes, and the semi-infinite substrate into only forward modes ($\mathbf{a}_{\rm t}$), yields to 
\begin{equation}
    \mathbf{W}_s^{+}\mathbf{a}_{\rm t} = \mathbf{P} \left( \mathbf{W}_0^{+}\mathbf{a}_{\rm in} + \mathbf{W}_0^{-}\mathbf{a}_{\rm r} \right). \label{eq:tmm_boundary}
\end{equation}
Solving Eq.~\ref{eq:tmm_boundary} directly provides the exact multilayer Jones reflection ($\mathbf{R}$) and transmission ($\mathbf{T}$) matrices:

$$\mathbf{R}= \begin{pmatrix} r_{pp} & r_{ps}\\ r_{sp} & r_{ss} \end{pmatrix}, \qquad \mathbf{T}= \begin{pmatrix} t_{pp} & t_{ps}\\ t_{sp} & t_{ss} \end{pmatrix}.$$

The magneto-optical effects discussed in the main text are extracted directly from the components of $\mathbf{R}$ and $\mathbf{T}$, completely replacing the single-interface Fresnel coefficients with the full multilayer interference amplitudes.

For the Faraday effect at normal incidence, the complex transmission amplitudes $t_{+}$ and $t_{-}$ for the circularly polarized eigenmodes yield the complex Faraday angle as
\begin{equation}
    \Phi_F^{\mathrm{TMM}} = \frac{i}{2} \ln\left(\frac{t_+}{t_-}\right). \label{eq:tmm_faraday}
\end{equation}
For a single forward-propagating pass through a homogeneous layer ($t_{\pm}\propto\exp(i k_0 n_{\pm}d)$), Eq.~\ref{eq:tmm_faraday} cleanly reduces to the bulk expression $-\frac{\pi d}{\lambda}(n_+-n_-)$ used in the main text.

For the Cotton--Mouton effect, the magnetic linear birefringence and dichroism are identically preserved using the linear transmission amplitudes
\begin{equation}
    \Phi_{cm}^{\mathrm{TMM}} = -\frac{1}{2} \ln\left(\frac{t_s}{t_p}\right).
\end{equation}
In the single-pass limit,
$t_s/t_p=\exp[i k_0d(n_s-n_p)]$, and therefore $\Phi_{cm}^{\mathrm{TMM}}
\rightarrow
-\frac{i\pi d}{\lambda}(n_s-n_p),$
which recovers Eq.~\ref{eq:Cotton-Mouton}. The finite-thickness formulation therefore preserves the same definition of magnetic linear birefringence and dichroism while adding the interface and interference contributions.

For the Polar, Transverse, and Longitudinal Kerr effects, no new structural formulas are required. The complex Kerr angles ($\Phi_p$, $\Phi_K^{s,p}$) and transverse reflectances ($\delta_{pp}$, $\delta_{ss}$) are calculated using the exact identical equations presented in the main text. The sole modification is substituting the single-interface Fresnel reflection coefficients with the corresponding matrix elements from the total TMM reflection matrix $\mathbf{R}$ ($r_{pp}$, $r_{ss}$, $r_{ps}$, $r_{sp}$). This seamlessly extends the polarization-selective MD response to finite thin-film structures.

The single-interface approximation used in the main text is only physically valid when the contribution of the remote BiYIG/substrate interface is negligible. For a mode with an imaginary longitudinal wave vector $\mathrm{Im}(k_z) = k_0\kappa$, the attenuation of the round-trip amplitude dictates the minimum thickness $d$ required to suppress back-reflections to an acceptable fractional error limit $\delta$
\begin{equation}
    d \gtrsim \frac{\lambda_0}{4\pi\kappa} \ln\left(\frac{\vert{}r_{\mathrm{back}}\vert{}}{\delta}\right).
\end{equation}

For a $1\%$ error limit ($\delta = 0.01$) assuming high back-reflection ($\vert{}r_{\mathrm{back}}\vert{}\le1$), this requires $d \gtrsim 0.37\frac{\lambda_0}{\kappa}$. In highly transparent or ultra-thin films where this condition fails, the coherent Fabry--Pérot interference is inescapable, and the finite-layer TMM must be used to accurately interpret phase-sensitive Kerr or Faraday measurements.

		\end{widetext}		

		\bibliography{main}

@article{krumme1989optical,
	title={Optical recording aspects of rf magnetron-sputtered iron-garnet films},
	author={Krumme, J-P and Doormann, V and Hansen, P and Baumgart, H and Petruzzello, J and Viegers, MPA},
	journal={Journal of applied physics},
	volume={66},
	number={9},
	pages={4393--4407},
	year={1989},
	publisher={American Institute of Physics}
}

@article{vojna2020faraday,
	title={Faraday rotation of Dy2O3, CeF3 and Y3Fe5O12 at the mid-infrared wavelengths},
	author={Vojna, David and Slez{\'a}k, Ond{\v{r}}ej and Yasuhara, Ryo and Furuse, Hiroaki and Lucianetti, Antonio and Mocek, Tom{\'a}{\v{s}}},
	journal={Materials},
	volume={13},
	number={23},
	pages={5324},
	year={2020},
	publisher={MDPI}
}

@article{walowski2016perspective,
  author = {Walowski, Jakob and M{\"u}nzenberg, Markus},
  title = {Perspective: Ultrafast magnetism and THz spintronics},
  journal = {Journal of Applied Physics},
  volume = {120},
  pages = {140901},
  year = {2016},
  doi = {10.1063/1.4958846}
}

@article{blank2021thz,
  author = {Blank, Thomas G. H. and Grishunin, Kirill A. and Mashkovich, Evgeny A. and Logunov, M. V. and Zvezdin, Andrei K. and Kimel, Alexey V.},
  title = {THz-Scale Field-Induced Spin Dynamics in Ferrimagnetic Iron Garnets},
  journal = {Physical Review Letters},
  volume = {127},
  pages = {037203},
  year = {2021},
  doi = {10.1103/PhysRevLett.127.037203}
}

@article{blank2023rectification,
  author = {Blank, Thomas G. H. and Mashkovich, Evgeny A. and Grishunin, Kirill A. and Schippers, C. F. and Logunov, M. V. and Koopmans, Bert and Zvezdin, Andrei K. and Kimel, Alexey V.},
  title = {Effective rectification of terahertz electromagnetic fields in a ferrimagnetic iron garnet},
  journal = {Physical Review B},
  volume = {108},
  pages = {094439},
  year = {2023},
  doi = {10.1103/PhysRevB.108.094439}
}

@article{hsu2020observation,
  author = {Hsu, Wei-Hung and Shen, Ka and Fujii, Yasuhiro and Koreeda, Akitoshi and Satoh, Takuya},
  title = {Observation of terahertz magnon of Kaplan-Kittel exchange resonance in yttrium-iron garnet by Raman spectroscopy},
  journal = {Physical Review B},
  volume = {102},
  pages = {174432},
  year = {2020},
  doi = {10.1103/PhysRevB.102.174432}
}

@article{kaplan1953exchange,
  author = {Kaplan, J. and Kittel, Charles},
  title = {Exchange Frequency Electron Spin Resonance in Ferrites},
  journal = {The Journal of Chemical Physics},
  volume = {21},
  pages = {760--761},
  year = {1953},
  doi = {10.1063/1.1699018}
}

@article{princep2017full,
  author = {Princep, Andrew J. and Ewings, Russell A. and Ward, Simon and Toth, Sandor and Dubs, Carsten and Prabhakaran, Dharmalingam and Boothroyd, Andrew T.},
  title = {The full magnon spectrum of yttrium iron garnet},
  journal = {npj Quantum Materials},
  volume = {2},
  pages = {63},
  year = {2017},
  doi = {10.1038/s41535-017-0067-y}
}

@article{shamoto2018neutron,
  author = {Shamoto, Shin-ichiro and Ito, Takahiro U. and Onishi, Hiroki and Yamauchi, Yoshinori and Inamura, Yasuhiro and Matsuura, Masahiro and Akatsu, Masato and Kodama, Kenji and Nakao, Atsushi and Moyoshi, T. and Munakata, Koji and Ohhara, Takenao and Nakamura, Masaaki and Ohira-Kawamura, Seiko and Nemoto, Yu and Shibata, Kenji},
  title = {Neutron scattering study of yttrium iron garnet},
  journal = {Physical Review B},
  volume = {97},
  pages = {054429},
  year = {2018},
  doi = {10.1103/PhysRevB.97.054429}
}

@article{ozgur2009microwave,
  author = {Ozgur, Umit and Alivov, Yahya and Morkoc, Hadis},
  title = {Microwave ferrites, part 1: Fundamental properties},
  journal = {Journal of Materials Science: Materials in Electronics},
  volume = {20},
  pages = {789--834},
  year = {2009},
  doi = {10.1007/s10854-009-9923-2}
}

@article{wettling1976magnetooptics,
  author = {Wettling, Wolfram},
  title = {Magneto-optics of ferrites},
  journal = {Journal of Magnetism and Magnetic Materials},
  volume = {3},
  pages = {147--160},
  year = {1976},
  doi = {10.1016/0304-8853(76)90026-3}
}

@article{freiser1968survey,
  author = {Freiser, Michael J.},
  title = {A Survey of Magnetooptic Effects},
  journal = {IEEE Transactions on Magnetics},
  volume = {4},
  pages = {152--161},
  year = {1968},
  doi = {10.1109/TMAG.1968.1066210}
}

@article{pershan1967magnetooptical,
  author = {Pershan, Peter S.},
  title = {Magneto-optical effects},
  journal = {Journal of Applied Physics},
  volume = {38},
  pages = {1482--1490},
  year = {1967}
}

@book{zvezdin1997modern,
  author = {Zvezdin, Andrei K. and Kotov, Vladimir A.},
  title = {Modern Magnetooptics and Magnetooptical Materials},
  publisher = {CRC Press},
  year = {1997},
  address = {Boca Raton, FL},
  isbn = {9781420050844}
}

@article{krinchik1959tensors,
  author = {Krinchik, G. S. and Chetkin, M. V.},
  title = {The Problem of Determining the Dielectric Permittivity and Magnetic Permeability Tensors of a Medium},
  journal = {Soviet Physics JETP},
  volume = {36},
  pages = {1368--1369},
  year = {1959}
}

@article{krinchik1960faraday,
  author = {Krinchik, G. S. and Chetkin, M. V.},
  title = {Faraday Effect in Yttrium Garnet at Infrared Frequencies},
  journal = {Soviet Physics JETP},
  volume = {11},
  pages = {1184--1186},
  year = {1960}
}

@article{krinchik1961infrared,
  author = {Krinchik, G. S. and Chetkin, M. V.},
  title = {Magneto-Optical Properties of Garnet Ferrites in the Infrared Region},
  journal = {Soviet Physics JETP},
  volume = {13},
  pages = {509--511},
  year = {1961}
}

@article{krinchik1962exchange,
  author = {Krinchik, G. S. and Chetkin, M. V.},
  title = {Exchange Interaction and Magneto-Optical Effects in Ferrite Garnets},
  journal = {Soviet Physics JETP},
  volume = {14},
  pages = {485--490},
  year = {1962}
}

@article{crossley1969faraday,
  author = {Crossley, W. A. and Cooper, R. W. and Page, J. L. and van Stapele, R. P.},
  title = {Faraday Rotation in Rare-Earth Iron Garnets},
  journal = {Physical Review},
  volume = {181},
  pages = {896--904},
  year = {1969},
  doi = {10.1103/PhysRev.181.896}
}

@article{takeuchi1973faraday,
  author = {Takeuchi, Hiroshi and Ito, Shin-ichi and Mikami, Itsuo and Taniguchi, Satoshi},
  title = {Faraday rotation and optical absorption of a single crystal of bismuth-substituted gadolinium iron garnet},
  journal = {Journal of Applied Physics},
  volume = {44},
  pages = {4789--4790},
  year = {1973},
  doi = {10.1063/1.1662042}
}

@article{wittekoek1975magnetooptic,
  author = {Wittekoek, S. and Popma, T. J. A. and Robertson, J. M. and Bongers, P. F.},
  title = {Magneto-optic spectra and the dielectric tensor elements of bismuth-substituted iron garnets at photon energies between 2.2-5.2 eV},
  journal = {Physical Review B},
  volume = {12},
  pages = {2777--2784},
  year = {1975},
  doi = {10.1103/PhysRevB.12.2777}
}

@article{hansen1984bismuth,
  author = {Hansen, P. and Witter, K. and Tolksdorf, W.},
  title = {Magnetic and magneto-optic properties of bismuth- and aluminum-substituted iron garnet films},
  journal = {Journal of Applied Physics},
  volume = {55},
  pages = {1052--1061},
  year = {1984}
}

@article{hansen1984review,
  author = {Hansen, P. and Krumme, J.-P.},
  title = {Magnetic and magneto-optical properties of garnet films},
  journal = {Thin Solid Films},
  volume = {114},
  pages = {69--107},
  year = {1984},
  doi = {10.1016/0040-6090(84)90337-7}
}

@article{booth1984garnet,
  author = {Booth, R. C. and White, E. A. D.},
  title = {Magneto-optic properties of rare earth iron garnet crystals in the wavelength range 1.1-1.7 micron and their use in device fabrication},
  journal = {Journal of Physics D: Applied Physics},
  volume = {17},
  pages = {579--587},
  year = {1984},
  doi = {10.1088/0022-3727/17/3/015}
}

@article{kang2012farinfrared,
  author = {Kang, T. D. and Standard, E. C. and Rogers, P. D. and Ahn, K. H. and Sirenko, A. A. and Dubroka, A. and Bernhard, C. and Park, S. and Choi, Y. J. and Cheong, S.-W.},
  title = {Far-infrared spectra of the magnetic exchange resonances and optical phonons and their connection to magnetic and dielectric properties of Dy3Fe5O12 garnet},
  journal = {Physical Review B},
  volume = {86},
  pages = {144112},
  year = {2012},
  doi = {10.1103/PhysRevB.86.144112}
}

@article{gribova2024inverse,
  author = {Gribova, N. I. and Berzhansky, V. N. and Polulyakh, S. N. and Belotelov, V. I.},
  title = {Inverse Faraday Effect in Ferrite--Garnet Films in the Near-Infrared Range},
  journal = {JETP Letters},
  volume = {120},
  pages = {183--189},
  year = {2024},
  doi = {10.1134/S0021364024601684}
}

@article{moskvin2022charge,
  author = {Moskvin, Alexander S.},
  title = {Charge Transfer Transitions and Circular Magnetooptics in Ferrites},
  journal = {Magnetochemistry},
  volume = {8},
  number = {8},
  pages = {81},
  year = {2022},
  doi = {10.3390/magnetochemistry8080081}
}

@article{atkinson1992sign,
  author = {Atkinson, R. and Lissberger, P.},
  title = {Sign conventions in magneto-optical calculations and measurements},
  journal = {Applied Optics},
  volume = {31},
  pages = {6076--6081},
  year = {1992},
  doi = {10.1364/AO.31.006076}
}

@article{wangness1954susceptibility,
  author = {Wangsness, R. K.},
  title = {Susceptibility Tensor and the Faraday Effect in Ferrimagnets},
  journal = {Physical Review},
  volume = {95},
  pages = {339--345},
  year = {1954}
}

@article{yang2022gyromagnetic,
  author = {Yang, Weihao and Liu, Qing and Wang, Hanbin and Chen, Yiqin and Yang, Run and Xia, Shuang and Luo, Yi and Deng, Longjiang and Qin, Jun and Duan, Huigao and Bi, Lei},
  title = {Observation of optical gyromagnetic properties in a magneto-plasmonic metamaterial},
  journal = {Nature Communications},
  volume = {13},
  pages = {1719},
  year = {2022},
  doi = {10.1038/s41467-022-29452-9}
}

@article{smolensky1978birefringence,
  author  = {Smolensky, G. A. and Pisarev, R. V. and Krichevtsov, B. B.},
  title   = {Magnetic birefringence and its relation to the sublattices in rare-earth iron garnets},
  journal = {Materials Research Bulletin},
  volume  = {13},
  number  = {12},
  pages   = {1441--1446},
  year    = {1978},
  doi     = {10.1016/0025-5408(78)90137-X}
}

@article{ando1987birefringence,
  author  = {Ando, K. and Le Gall, H. and Desvignes, J. M.},
  title   = {Linear birefringence in Bi-substituted iron garnet crystals},
  journal = {Journal of the Magnetics Society of Japan},
  volume  = {11},
  number  = {S_1_ISMO},
  pages   = {S1-295--S1-298},
  year    = {1987},
  doi     = {10.3379/jmsjmag.11.S1_295}
}

@article{sato1989cotton,
  author  = {Sato, K. and Aman, Y. and Tamanoi, K. and Saito, T. and Shinagawa, K. and Tsushima, T.},
  title   = {Cotton-Mouton Spectrum in Cobalt-Substituted Magnetic Garnet},
  journal = {Journal of the Magnetics Society of Japan},
  volume  = {13},
  number  = {2},
  pages   = {157--162},
  year    = {1989},
  doi     = {10.3379/jmsjmag.13.157}
}

@article{baierl2016thz,
  author  = {Baierl, S. and Mentink, J. H. and Hohenleutner, M. and Braun, L. and Do, T.-M. and Lange, C. and Sell, A. and Fiebig, M. and Woltersdorf, G. and Kampfrath, T. and Huber, R.},
  title   = {Terahertz-Driven Nonlinear Spin Response of Antiferromagnetic Nickel Oxide},
  journal = {Physical Review Letters},
  volume  = {117},
  pages   = {197201},
  year    = {2016},
  doi     = {10.1103/PhysRevLett.117.197201}
}

@article{sato2022fundamentals,
  author  = {Sato, Katsuaki and Ishibashi, Takayuki},
  title   = {Fundamentals of Magneto-Optical Spectroscopy},
  journal = {Frontiers in Physics},
  volume  = {10},
  pages   = {946515},
  year    = {2022},
  doi     = {10.3389/fphy.2022.946515}
}

@article{berreman1972,
  author  = {Berreman, Dwight W.},
  title   = {Optics in Stratified and Anisotropic Media: 4 x 4-Matrix Formulation},
  journal = {Journal of the Optical Society of America},
  volume  = {62},
  number  = {4},
  pages   = {502--510},
  year    = {1972},
  doi     = {10.1364/JOSA.62.000502}
}

@article{zak1990universal,
  author  = {Zak, J. and Moog, E. R. and Bader, S. D.},
  title   = {Universal approach to magneto-optics},
  journal = {Journal of Magnetism and Magnetic Materials},
  volume  = {89},
  number  = {1--2},
  pages   = {107--123},
  year    = {1990},
  doi     = {10.1016/0304-8853(90)90713-Z}
}

@article{zak1991multilayer,
  author  = {Zak, J. and Moog, E. R. and Liu, C. and Bader, S. D.},
  title   = {Magneto-optics of multilayers with arbitrary magnetization directions},
  journal = {Physical Review B},
  volume  = {43},
  number  = {8},
  pages   = {6423--6429},
  year    = {1991},
  doi     = {10.1103/PhysRevB.43.6423}
}

@article{atkinson1993firstorder,
  author  = {Atkinson, R. and Lissberger, P. H.},
  title   = {Correct formulation of first-order magneto-optical effects in multilayer thin films in terms of characteristic matrices and derivation of a related superposition principle},
  journal = {Journal of Magnetism and Magnetic Materials},
  volume  = {118},
  number  = {1--2},
  pages   = {271--277},
  year    = {1993},
  doi     = {10.1016/0304-8853(93)90183-3}
}

@article{rogers2011mueller,
  author  = {Rogers, P. D. and Kang, T. D. and Zhou, T. and Kotelyanskii, M. and Sirenko, A. A.},
  title   = {Mueller matrices for anisotropic metamaterials generated using 4 x 4 matrix formalism},
  journal = {Thin Solid Films},
  volume  = {519},
  number  = {9},
  pages   = {2668--2673},
  year    = {2011},
  doi     = {10.1016/j.tsf.2010.12.066}
}

@article{zamani2016bigyrotropic,
  author  = {Zamani, Mehdi and Nezhad Hajesmaeili, Hamidreza and Zandi, Mohammad Hossein},
  title   = {Transfer matrix method-based approach to study the bi-gyrotropic magnetic materials},
  journal = {Optical Materials},
  volume  = {58},
  pages   = {38--45},
  year    = {2016},
  doi     = {10.1016/j.optmat.2016.03.044}
}


\end{document}